\documentclass[%
 reprint,
superscriptaddress,
nofootinbib,
 amsmath,amssymb,
 aps,
pra,
]{revtex4-2}
\usepackage{physics}
\usepackage{bbold}
\usepackage{amsmath}
\usepackage{latexsym}
\usepackage{graphicx}
\usepackage{amsthm}
\usepackage{hyperref}
\usepackage{lipsum}
\usepackage[english]{babel}
\usepackage{slashed}
\usepackage[compat=1.1.0]{tikz-feynman}
\usepackage{appendix}
\usepackage{subfig}
\usepackage{multirow}
\usepackage{mathtools}
\usepackage{subfig}
\usepackage{pgfplots}
\pgfplotsset{compat=1.15}
\usepackage{mathrsfs}
\usetikzlibrary{arrows}
\definecolor{ffffff}{rgb}{1,1,1}
\definecolor{qqqqff}{rgb}{0,0,1}
\definecolor{qqwuqq}{rgb}{0,0.39215686274509803,0}
\definecolor{zzttqq}{rgb}{0.6,0.2,0}
\definecolor{ttqqqq}{rgb}{0.2,0,0}
\definecolor{qqttqq}{rgb}{0,0.2,0}
\definecolor{qqccqq}{rgb}{0,0.8,0}
\usepackage{tikz-cd}
\usepackage{amsmath}
\usepackage{amsbsy}

\newcommand{\adr}[1]{{\color{brown} #1}}
\usepackage{ragged2e}
\usepackage{comment}

\theoremstyle{plain}

\tikzset{
pattern size/.store in=\mcSize, 
pattern size = 5pt,
pattern thickness/.store in=\mcThickness, 
pattern thickness = 0.3pt,
pattern radius/.store in=\mcRadius, 
pattern radius = 1pt}
\makeatletter
\pgfutil@ifundefined{pgf@pattern@name@_b8864pxwu}{
\pgfdeclarepatternformonly[\mcThickness,\mcSize]{_b8864pxwu}
{\pgfqpoint{0pt}{0pt}}
{\pgfpoint{\mcSize+\mcThickness}{\mcSize+\mcThickness}}
{\pgfpoint{\mcSize}{\mcSize}}
{
\pgfsetcolor{\tikz@pattern@color}
\pgfsetlinewidth{\mcThickness}
\pgfpathmoveto{\pgfqpoint{0pt}{0pt}}
\pgfpathlineto{\pgfpoint{\mcSize+\mcThickness}{\mcSize+\mcThickness}}
\pgfusepath{stroke}
}}
\makeatother

\tikzset{
pattern size/.store in=\mcSize, 
pattern size = 5pt,
pattern thickness/.store in=\mcThickness, 
pattern thickness = 0.3pt,
pattern radius/.store in=\mcRadius, 
pattern radius = 1pt}
\makeatletter
\pgfutil@ifundefined{pgf@pattern@name@_f6j18z8qd}{
\pgfdeclarepatternformonly[\mcThickness,\mcSize]{_f6j18z8qd}
{\pgfqpoint{0pt}{0pt}}
{\pgfpoint{\mcSize+\mcThickness}{\mcSize+\mcThickness}}
{\pgfpoint{\mcSize}{\mcSize}}
{
\pgfsetcolor{\tikz@pattern@color}
\pgfsetlinewidth{\mcThickness}
\pgfpathmoveto{\pgfqpoint{0pt}{0pt}}
\pgfpathlineto{\pgfpoint{\mcSize+\mcThickness}{\mcSize+\mcThickness}}
\pgfusepath{stroke}
}}
\makeatother
\tikzset{every picture/.style={line width=0.75pt}} 

\usepackage[bottom]{footmisc}

\begin{document}

\title{The thermodynamics of readout devices and semiclassical gravity}
\author{Samuel Fedida}
\affiliation{Centre for Quantum Information and Foundations, DAMTP, Centre for Mathematical Sciences, University of Cambridge, Wilberforce Road, Cambridge CB3 0WA, UK}
\author{Adrian Kent}
\affiliation{Centre for Quantum Information and Foundations, DAMTP, Centre for Mathematical Sciences, University of Cambridge, Wilberforce Road, Cambridge CB3 0WA, UK}
\affiliation{Perimeter Institute for Theoretical
	Physics, 31 Caroline Street North, Waterloo, ON N2L 2Y5, Canada.}
\date{\today}

\begin{abstract}
    We analyse the common claim that nonlinear modifications of quantum theory necessarily violate the second law of thermodynamics. We focus on hypothetical extensions of quantum theory that contain readout devices. These black boxes provide a classical description of quantum states without perturbing them.  They allow quantum state cloning, though in a way consistent with the relativistic no-signalling principle. We review the existence of such devices in the context of M{\o}ller-Rosenfeld semiclassical gravity, which postulates that the gravitational field remains classical and is sourced by the expectation value of a quantum energy-momentum tensor. We show that the definition of information in the models examined in this paper deviates from that given by von Neumann entropy, and that claims of second law violations based on the distinguishability of non-orthogonal states or on violations of uncertainty principles fail to hold in such theories.
\end{abstract}

\maketitle

\section{Introduction}

In standard quantum mechanics, the act of measurement extracts classical information from a quantum state whilst disturbing the latter, with bounds on the tradeoff between information obtained and disturbance.  Nonlinear modifications to the Schrödinger equation can, in principle, violate these bounds.  For example, quantum theory can be consistently extended by introducing readout devices \cite{Kent2021b}, which are idealized black boxes capable of extracting certain properties of a quantum state without inducing wavefunction collapse or otherwise altering the state. 
One theoretical motivation for considering such extensions is that M{\o}ller-Rosenfeld semiclassical gravity effectively introduces finite-precision versions of such devices.

M{\o}ller-Rosenfeld semiclassical gravity is based on the hypothesis that spacetime is classical and couples to quantum matter through the expectation value of the energy-momentum tensor of matter. The semiclassical field equations then read
\begin{equation}
    \label{Semiclassical Field Equations}
    G_{\mu \nu} = \frac{8\pi G}{c^4} \expval{\hat{T}_{\mu\nu}}
\end{equation}
It is not evident that eqn. \eqref{Semiclassical Field Equations} defines a 
complete consistent theory.
If it does, it is inconsistent with
astronomical observation, assuming that there is a universal matter wave function in a superposition significantly different from the observed matter state.    It is also inconsistent with  table-top experiments \cite{Page1981}.  
However, neither of these rule out \eqref{Semiclassical Field Equations} being valid as an effective limit \cite{WilsonGerow2022,Kuo1993}
within a restricted domain, or as part of 
a theory that modifies standard quantum dynamics (e.g. \cite{Tilloy2016,tilloy_binding_2018}).   

One argument \cite{Gisin1990} often made against nonlinear theories such as semiclassical gravity is that they necessarily introduce superluminal signalling. 
However, there is a quite general construction 
that can be used in such theories to avoid superluminal signalling \cite{Kent2005}. 

It has also been argued \cite{Masanes2025} that the measurement postulates
of quantum theory follow from other essential
defining postulates of quantum theory, which
nonlinear versions of quantum theory, which necessarily introduce nonstandard measurements, must thus violate.
These arguments were originally framed with several tacit
assumptions \cite{Kent2025, Stacey2024}.
When made explicit, careful analysis \cite{Masanes2025,Kent2025,Stacey2024} shows that 
nonlinear theories of the type discussed in \cite{Kent2005,Kent2025} extend quantum theory by
introducing new features in new regimes, without
breaking any principle required for the success
of quantum theory in experiments to date.

It is also worth noting that, while a semiclassical gravity theory
would violate the quantum no-signalling principle
(which we stress is distinct from the no-superluminal
signalling principle), it also provides an obvious
potential mechanism for the relevant light-speed signals,
namely the gravitational degrees of freedom.   
That said, one might also ask, even in the absence of such a 
mechanism, whether we should be so surprised 
by a physical theory in which 
actions on one subsystem have causal effects on
another, given that quantum theory itself includes
Bell non-local correlations.  We think that, for 
example, Einstein, Podolsky and Rosen \cite{Einstein1935} might have found the former less problematic than the latter. 

From the perspective of axiomatic relativistic quantum field theory, it also appears that relativistic no-signalling should be understood as an epistemic principle rather than an ontological one, i.e., taking interactions to be local in some ontological sense appears not only unnecessary but also unrealistic.
For example, Wightman's theorem \cite{wightman_theorie_1964,Halvorson2006col} states that imposing pointwise microcausality (i.e. $\comm{\hat{\phi}(x)}{\hat{\phi}(y)} = 0$ whenever $x$ and $y$ are spacelike separated points) is too strong for theories of translation-covariant quantum fields which satisfy the spectrum condition (the Hamiltonian is bounded from below) provided there exists a unique translation-invariant vacuum state: it implies that all $n$-point vacuum expectation values are constant over spacetime. Instead, weaker notions such as Einstein causality (measurements conducted over spacelike-separated regions commute) are usually assumed.

These various points give some extra motivation, if any were needed, to the active ongoing experimental efforts (e.g. \cite{Bose2017,Marletto2017,Howl2021,Kent2021a,Kent2021c}) to determine whether gravity is quantum, since models involving semiclassical gravity in some relevant regime give a concrete alternative that the relevant experiments can distinguish from quantum gravity. 

However, there are other independent arguments made against nonlinear theories such as semiclassical gravity, arising from claims that they violate the fundamental
principles of thermodynamics, and in particular the second law. 
These arguments go back to von Neumann \cite{VonNeumann1955} and were 
elaborated by Peres \cite{Peres1993}. More recently, Hänggi and Wehner showed \cite{Hanggi2013} that, under certain assumptions, violations of some versions of the uncertainty principle imply a violation of the second law of thermodynamics. Nonlinear quantum theories can have \cite{Mielnik1979,Galley2023} a classical state space in which non-orthogonal states can be perfectly distinguished, as readout devices \cite{Kent2005} illustrate,
and so evidently 
violate standard uncertainty principles.
The arguments of \cite{VonNeumann1955,Peres1993,Hanggi2013} thus suggest that
they indeed violate the second law and are implausible for this reason.
Such lines of thought have been taken (e.g. \cite{Terno2024}) to disfavour classical-quantum hybrid models based on nonlinear dynamics.

These arguments are also significant because, regardless of its fundamental status, semiclassical gravity is often taken as a reasonable approximation in many contexts, including for the Hawking radiation of black holes and for perturbation theory in inflationary cosmology, where one considers quantum field theory (QFT) in curved but classical spacetime.
In these contexts, understanding the thermodynamic and information theoretic properties of the relevant systems is key. 

It should be noted here that the predictions of M{\o}ller-Rosenfeld semiclassical gravity deviate in these regimes from those of the semiclassical limit 
of quantum gravity \cite{Fedida2025}.   Our discussion in the present paper is relevant 
to the former.   It might -- depending on the reader's perspective and
notwithstanding our conclusions -- reinforce
the view that the correct approximation is given by the latter. This is important to highlight to distinguish nuances in the literature. For example, in black hole complementarity \cite{Susskind1993}, deriving a form of quantum state cloning from the assumption that semiclassical gravitational equations hold alongside some fundamental unitarity is not the same as deriving cloning from assuming M{\o}ller-Rosenfeld semiclassical gravity. 
The latter is a straightforward consequence of the fact that M{\o}ller-Rosenfeld semiclassical gravity allows the construction of quantum state readout devices \cite{Kent2021b,Fedida2025}.   

In this paper, we consider whether semiclassical gravity and, more generally, nonlinear extensions of quantum theory with readout devices (which we review in the next section) are consistent with thermodynamics.   We show that a careful treatment need not contradict appropriately formulated thermodynamic laws.

Interestingly, a second law of thermodynamics can be justified for generalised probabilistic theories \cite{Krumm2017} 
and for models combining classical and quantum mechanics \cite{Layton2025}.   
The models analysed in \cite{Krumm2017} satisfy two postulates, which do not typically hold in the class of models we consider here. 
Ref. \cite{Layton2025} analyses models with a specific stochastic coupling between classical and quantum degrees of freedom. 
The models discussed in Refs. \cite{Krumm2017,Layton2025} are linear; we consider nonlinear extensions of quantum theory.

\section{Readout devices}

We begin by reviewing the definitions of and some motivations
for introducing \textit{readout devices} \cite{Kent2005,Kent2021b,Kent2021c,Kent2025}. 
These are hypothetical idealised devices that give us classical information about a pure quantum state $\ket{\psi} \in \mathcal{H}_1 \otimes ... \otimes \mathcal{H}_N$ of a set of N systems.
The devices we consider cannot be constructed using standard quantum measurements, so considering them implies considering extensions of standard quantum theory.   

We will take the relevant $\mathcal{H}_i$, namely those whose degrees of freedom the devices give information about, to be finite-dimensional Hilbert spaces. Finite precision (FP) models of such devices may also be described.   In one simple model, an FP readout device takes a positive integer $l$ as additional input, and outputs the associated information to $l$ digits of binary precision.   When the input is a pure state, the action of the FP readout device may be defined either on the ray in Hilbert space or on the corresponding density matrix: the finite precision versions of these are approximately but not generally exactly equivalent.  

We define a \textit{state readout device} (RD) to act on subsystems: given subsystems $1, \cdots,i \leq n$ of the state $[\psi]$ of a system $S$ it outputs a classical description (in some given basis) of $\rho_{1 \cdots i} = \Tr_{i+1, \cdots,N}(\ket{\psi}\bra{\psi}) \in \mathcal{L}(\mathcal{H}_1 \otimes \cdots \otimes \mathcal{H}_i)$ where $\mathcal{L}(\mathcal{H}_j)$ denotes the space of linear operators on $\mathcal{H}_j$ for $j \in \{1, \cdots, n\}$.\footnote{In Ref. \cite{Kent2025} an RD is defined to act on subsystem $1$.   The definition here is equivalent, up to redefinition of subsystems, but more convenient for our discussion.} That is,

\begin{widetext}
    \begin{equation}
    \label{semiclassical RD}
    l, \text{basis}, \rho_{1 \cdots i} \xlongrightarrow{input} \boxed{\text{(FP)RD}} \xlongrightarrow{output} \begin{cases}
        a_{11} = 0.n_{11}^{(1)}n_{11}^{(2)}...n_{11}^{(l)} \\
        a_{12} = 0.n_{12}^{(1)}n_{12}^{(2)}...n_{12}^{(l)} + i  0.p_{12}^{(1)}p_{12}^{(2)}...p_{12}^{(l)}
         \\
        \vdots \\ a_{k-1,k} = 0.n_{k-1,k}^{(1)}n_{k-1,k}^{(2)}...n_{k-1,k}^{(l)} + i0.p_{k-1,k}^{(1)}p_{k-1,k}^{(2)}...p_{k-1,k}^{(l)} \\
        a_{kk} = 0.n_{kk}^{(1)}n_{kk}^{(2)}...n_{kk}^{(l)} 
    \end{cases}
\end{equation}
\end{widetext}
where $k = \prod_{j=1}^i \dim \mathcal{H}_j$. The density matrix $\rho_{1,\cdots,i}$ is parametrised by $k^2-1$ real parameters, so in the finite precision case ($l < \infty$), this requires $ \approx (k^2-1) l$ bits of (classical) information storage.

A \textit{probability readout device} (PRD) outputs a classical description of the diagonal elements (Born probabilities) of $\rho_{1,\cdots,i}$ in some given basis:
\begin{widetext}
    \begin{equation}
    \label{semiclassical PRD}
    l, \text{basis}, \rho_{1 \cdots i} \xlongrightarrow{input} \boxed{\text{(FP)PRD}} \xlongrightarrow{output} \begin{cases}
        \abs{a_{11}}^2 = 0.n_{11}^{(1)}n_{11}^{(2)}...n_{11}^{(l)} \\
        \abs{a_{22}}^2 = 0.n_{22}^{(1)}n_{22}^{(2)}...n_{22}^{(l)}
         \\
        \vdots \\ 
        \abs{a_{kk}}^2 = 0.n_{kk}^{(1)}n_{kk}^{(2)}...n_{kk}^{(l)} 
    \end{cases} \quad .
\end{equation}
\end{widetext}
In the finite-precision case, this requires $\approx (k-1)l$ bits of (classical) information storage.
Here, note that we only have information about the probabilities, not the 
amplitudes or their relative phases.   Using quantum state tomography, 
the amplitudes (up to an overall phase) can be obtained by repeated use of a PRD, but the information-theoretic analysis is more involved and not relevant for our discussion here.

An \textit{expectation value readout device} (ERD) takes as input some hermitian observable A defined on systems $1 \cdots i$, and outputs the expectation value $\Tr(A \rho_{1 \cdots i})$. In the infinite precision case, the expectation values can be obtained from PRDs plus post-processing when the PRDs work in the eigenbasis of the observable. In the finite precision case, this is approximately true for sufficiently large $l$, but the detailed propagation of the precision is more subtle and depends on the observable and post-processing.

Note that an infinite precision RD is more powerful than a perfect cloning device on a pure state with $i=N$,
assuming that perfect infinite precision classical and quantum operations are free resources.  
Reading $\ket{\Psi}$ gives classical information about that state that we may encode in a quantum state in some basis. That is, starting with a blank ancilla state $\ket{*}$, we can implement
\begin{equation}
    \ket{\Psi}\ket{*} \longrightarrow \boxed{\text{RD \& enc.}} \longrightarrow \ket{\Psi}\ket{\Psi} \, , 
\end{equation}
up to a global phase, so that one use of an infinite precision RD allows perfect cloning.
On the other hand, a perfect cloning device would need to be used uncountably infinitely often, using uncountably many ancillae (which goes beyond standard assumptions about the physically accessible states in quantum field theory) to generate the quantum information required to produce an infinite
precision classical description.  

Even if quantum theory is not fundamental, it seems hard to imagine that infinite precision state readout devices (or infinite precision versions of the other devices) could be operationally realised in nature, since the infinite precision outputs cannot be expressed in a finite time. These devices should rather be thought of as convenient mathematical idealisations. 

At first glance, it may also seem hard to motivate the hypothesis that anything resembling a FPRD could be found in nature, given that FPRDs define nonstandard extensions of standard quantum theory that violate quantum no-signalling and no-cloning, albeit without necessarily violating relativistic causality \cite{Kent2005}. However, the possibility that semiclassical gravity holds true in some regime provides an explicit example of a type of theory, considered for independent reasons, that would imply FPRDs \cite{Kent2021b,Fedida2025}.   More generally, nonlinear extensions of quantum theory generically imply some form of FPRD.   

One might take this as an argument against semiclassical gravity and other nonlinear theories.   However, to make a compelling argument one needs to show that extensions of quantum theory involving readout devices necessarily have incurable problems.   Perhaps they do, but our aim in this paper is to show that the thermodynamic arguments often cited in this context do not identify such problems.   

\subsection{An example: ERDs and PRDs in semiclassical gravity}

The standard assumption in the context of semiclassical gravity is that the gravitational field could be measured without affecting the quantum matter state.  
This would imply that a device measuring the classical gravitational field at any point would act as a finite-precision ERD.   
If this held true, then versions of FPPRDs would exist and could in principle be constructed \cite{Kent2021b} in the relevant regime, as we now review.
We can simplify the information theoretic analysis by looking at the nonrelativistic limit of the semiclassical Einstein field equations \eqref{Semiclassical Field Equations}, 
noting again that the construction of readout devices can be defined to avoid superluminal signalling \cite{Kent2005}. 

In this limit, we can define the semiclassical Newtonian potential to follow Poisson's equation
\begin{equation}
    \label{Semiclassical Newtonian potential}
    \Delta \Phi(\abs{\psi}^2;\mathbf{x}) = 4\pi G \expval{\hat{\rho}(\mathbf{x})}
\end{equation}
where $\hat{\rho}(x)$ is the mass density operator of the quantum matter, so that the Schrödinger equation has an extra term from the gravitational Hamiltonian \cite{Jones1995} and becomes the Schrödinger-Newton Equation (SNE) \cite{Ruffini1969}
\begin{equation}
    \label{Schrodinger-Newton equation}
    i\hbar \frac{\partial}{\partial t} \ket{\psi(t)} = \Big(\hat{H}_{matter} + \int \hat{\rho}(\mathbf{x}) \Phi(\abs{\psi}^2;\mathbf{x}) d^3 \mathbf{x}\Big)\ket{\psi(t)} \, .
\end{equation}
For a single particle of mass $m$ in a superposition of localised positions with quantum state 
\begin{equation}
    \ket{\psi} = a_0 \ket{\mathbf{x}_0} + a_1 \ket{\mathbf{x}_1} \, ,
\end{equation}
the semiclassical gravitational field at some point $\mathbf{y}$ can be found from Poisson's equation \eqref{Semiclassical Newtonian potential} and is given by
\begin{equation}
    \label{Semiclassical gravitational field}
    \Phi(\mathbf{y}) = -Gm\Big(\frac{\abs{a_0}^2}{\abs{\mathbf{x}_0-\mathbf{y}}} + \frac{\abs{a_1}^2}{\abs{\mathbf{x}_1-\mathbf{y}}}\Big) \, .
\end{equation}
Determining the classical gravitational field at some other point $\mathbf{y'}$ provides information on the state $\ket{\psi}$ by giving us an estimate of the $\abs{a_i}^2$ for $i=1,2$.
In fact, given that we can take $\ket{\psi}$ to be normalised, just one reading suffices. 
More generally, given (for simplicity) a normalised pure quantum state
\begin{equation}
    \label{General position state}
    \ket{\Psi} = \sum_{i=1}^k a_i \ket{\mathbf{x}_i}
\end{equation}
we can estimate the $| a_i |^2 $ 
from readings of the semiclassical gravitational field
\begin{equation}
    \Phi(\mathbf{y}_j) = -G m \sum_{i=1}^k \frac{\abs{a_i}^2}{\abs{\mathbf{x}_i-\mathbf{y}_j}}
\end{equation}
at  $k-1$ locations $\mathbf{y}_j$.
We expect these readings to be finite precision because experimental constraints,
including the uncertainty principle, prevent us both from specifying the $\mathbf{y_j}$ 
and from measuring $\Phi(\mathbf{y}_j)$ to 
infinite precision.  

Thus, semiclassical gravity allows us in principle to construct an (FP)PRD of the form
\begin{widetext}
    \begin{equation}
    \label{Semiclassical gravity FPRD}
    l, \text{basis}, \ket{\Psi} \xlongrightarrow{input} \boxed{\text{Semiclassical gravity (FP)PRD}} \xlongrightarrow{output} \begin{cases}
        \abs{a_1}^2 = 0.n_1^{(1)}n_1^{(2)}...n_1^{(l)} \\
        \abs{a_2}^2 = 0.n_2^{(1)}n_2^{(2)}...n_2^{(l)} \\
        \vdots \\
        \abs{a_k}^2 = 0.n_k^{(1)}n_k^{(2)}...n_k^{(l)}
    \end{cases} \quad .
\end{equation}
\end{widetext}

In summary, readings of classical gravitational fields would be equivalent to finite precision estimates of the 
expectation value of the field generated by the quantum state, i.e. to some form of finite precision ERD.  
Repeated readings of such ERDs would provide some form of finite precision PRD. The finite precision estimates provided would in general include noise arising from gravitational fields extrinsic to the system being measured. Thus, the basis and degree of precision would depend on the underlying theory and on the specific physical context, 
and the degree of precision may vary for different states and different components of a state. The definitions of finite precision devices above thus also need to be understood as mathematical idealisations of the expected behaviour of semiclassical gravity and similar models. Nonetheless they capture the essential features of non-standard measurements available within these models.  

As was shown in \cite{Barnum2007}, no-cloning holds for any generalised probabilistic theory (GPT) -- including quantum theory -- apart from classical probability theory. There is no contradiction here since the fact that semiclassical gravity implies readout devices means that it has classical features (see also discussions by \cite{Mielnik1979,Galley2023}).   This can be understood in the sense of the theory allowing perfect discrimination of non-orthogonal states. Moreover, the notion of quantum state cloning is closely related to the notion of conservation of quantum information. Again, there is no contradiction: we will see that the notion of information changes in the context of semiclassical gravity and, more broadly, in theories with readout devices.

Given these theoretical motivations, it is interesting to understand the thermodynamical behaviour of quantum-like theories that include readout devices. Indeed, several arguments in the literature can naively be interpreted as ruling such theories out on the basis of thermodynamical instability. In this paper, we address these and show this interpretation is not justified.  

\subsection{Entropy in readout device world}

Extensions of quantum theory allow different definitions of entropy that
reduce to von Neumann entropy in the relevant regimes \cite{Short2010}. One such is the so-called \emph{decomposition entropy}, defined for a $d$-dimensional system as 
\begin{equation}\label{decomposition entropy}
    S_{\text{dec}}(\rho) := \inf_{\substack{\{p_j,\sigma_j\} \\ \rho = \sum_{j=1}^d p_j \sigma_j}} H(\{p_1,...,p_d\})
\end{equation}
where $H$ is the Shannon entropy and the infimum is taken over decompositions into perfectly distinguishable pure states \cite{Hanggi2013}. 

Decomposition entropy is proven to reduce to the Shannon and von Neumann entropies in classical probability theory and quantum theory in \cite{Short2010}. Although interesting in discussions of entanglement or purity, decomposition entropy does not satisfy concavity nor subadditivity in general theories (e.g. in box world \cite{Short2010}), and has a less direct operational interpretation than the \emph{measurement entropy}, which we now review.

First, we discuss the notion of \emph{measurements} and, in particular, of \emph{fine-grained measurements} \cite{Short2010}. A measurement is a set
\begin{equation}
    \mathbf{e} = \left\{(j,e_j) \mid j \in J \, \& \sum_{j \in J} e_j = \mathbb{1} \right\}
\end{equation}
where $J$ is an indexing set (the set of outcomes of the measurement) and $e_j : \mathscr{D}(\mathcal{H}) \to [0,1]$ are maps, called \emph{effects}, such that $e_j(\rho)$ is the probability of obtaining outcome $j$ given state $\rho$, and $\sum_{j \in J} e_j(\rho) = 1$ for all $\rho \in \mathscr{D}(\mathcal{H})$. 

Some measurements on a system can be more informative than others. Consider two measurements $\mathbf{e}$ and $\mathbf{f}$ for which there exists a map $M : J_\mathbf{e} \to J_{\mathbf{f}}$ such that
\begin{equation}
    \label{eqn:coarse graining}
    f_{j'}=\sum_{j \in J_{\mathbf{e}} : M(j) = j'} e_j \qquad \forall j' \in J_\mathbf{f} \, .
\end{equation}
If $M$ is one-to-one then this is just a re-labelling of outcomes, otherwise $\mathbf{f}$ is a \emph{coarse-graining} of $\mathbf{e}$ and, conversely, $\mathbf{e}$ is said to be a \emph{refinement} of $\mathbf{f}$. A coarse-graining is said to be \emph{trivial} if
\begin{equation}
    e_j \propto f_{M(j)} \qquad \forall j \in J_{\mathbf{e}} \, ,
\end{equation}
in which case the measurement $\mathbf{e}$ is equivalent to that of $\mathbf{f}$, obtaining $j'$ and outputting a randomly selected $j$ satisfying $M(j) = j'$. Thus, the two measurements are equally informative about the state of the system. A measurement $\mathbf{e}$ is said to be \emph{fine-grained} if it has no non-trivial refinements. Fine-grained measurements are thus those measurements which cannot be refined to give more information about the state.

The \emph{measurement entropy} $\hat{H}$ of a state $\rho$ is the infimum of the Shannon entropy of the probabilities associated to effects of fine-grained measurements:
\begin{equation}\label{measurement entropy}
    \hat{H}(\rho) := \inf_{\{e_j\}} H(\{e_j( \rho) \}) \, .
\end{equation}
It has a clear information-theoretic and operational meaning: it is the infimal output uncertainty of fine-grained measurements on the system.   A coarse-graining of a fine-grained measurement can never give more information \cite{Short2010}.

The definition extends to uncountable $J$ by replacing sums by
integrals and using the Shannon entropy of the probability distribution function.  

Infinite-precision readout devices correspond to the collection of effects 
$\{e_{\rho} \}_{\rho \in \mathscr{D}(\mathcal{H})}$ where 
\begin{equation}
    e_{\rho}(\sigma) = \begin{cases}
        \delta_{\rho,\sigma} \qquad & \text{ if } \sigma \text{ is improper} \, , \\
        \sum_i p_i \delta_{\rho, \sigma_i} \qquad &\text{ if } \sigma = \{p_i,  \sigma_i\} \text{ is proper} \, .
    \end{cases}
\end{equation}

In the case of infinite precision readout devices, the measurement entropy is generally different to the decomposition entropy.
The ``best measurements" are given by the readout devices: a measurement that provides all the information about a state cannot be further (non-trivially) refined. 
In particular, the effects of standard quantum measurements can be reproduced by the output of an infinite-precision readout device together with that of an infinite-precision random number generator.
Formally, 
any quantum effect $f_j$ can be written as 
\begin{equation}\label{qeffrd}
    f_j = \int_{\rho  \in \mathscr{D}(\mathcal{H})} p(\rho) e_\rho d \rho \,
\end{equation}
where $p(\rho)$ are appropriate probability weights and 
$\int d \rho$ is defined so that for any effect $g$,
\begin{equation}
\left(\int g( \rho ) e_{\rho} d \rho  \right) (\sigma ) = g (\sigma) \, .
\end{equation} 
Taking eqn.~\eqref{qeffrd} as a definition of coarse-graining in this setting, standard quantum effects are coarse-grainings of infinite-precision readout devices.     

Thus, given any proper mixture $\rho_\text{proper}$ associated to an ensemble $\{p_i,\ket{\psi_i}\}$,
\begin{align}
    S_{RD}(\{p_i,\ket{\psi_i}\})  :=& \hat{H}(\rho_\text{proper})  \\ =& H(\{p_i\}) \nonumber \\ = & - \sum_i p_i \log(p_i) \nonumber , .
\end{align}
That is, the measurement entropy of an ensemble in readout device world is always the Shannon entropy associated to the preparation probabilities. However, for improper mixtures, if one works with infinite-precision readout devices, one can fully determine $\rho_\text{improper}$ with probability 1, since $e_{\rho}(\rho_\text{improper}) = \delta_{\rho, \rho_\text{improper}}$.   Hence $S_{RD}(\rho_{improper}) = 0$. 

Finite-precision readout devices can also be characterised in the language of effects and fine-grained measurements. In the basis $\{\ket{\chi_i}\} \subset \mathcal{H}$, the effects are $\{e_{s}\}_{s \in S_\chi(l)}$ where
\begin{widetext}
    \begin{equation}
    S_\chi(l) = \left\{\sum_{i,j} a_{ij} \ket{\chi_i}\bra{\chi_j} \mid   a_{ij} = 0.n_{ij}^{(1)} \cdots n_{ij}^{(l)} + i 0.p_{ij}^{(1)} \cdots p_{ij}^{(l)}\right\} 
\end{equation}
\end{widetext}
is the set of allowed finite precision outputs.    
Then 
\begin{equation}
    e_{s}(\sigma) = \begin{cases}
        1 \quad & \text{ if } \sigma \approx s \text{ is improper} \, , \\
         p_i \quad &\text{ if } \sigma = \{p_i,  \sigma_i\}  \text{ is proper and } \sigma_i \approx s \, , \\
        0 \quad &\text{ otherwise}.
    \end{cases}
\end{equation}
where $\approx$ is to $l$ digits of precision. Thus FPRDs can determine an approximation of $\rho_\text{improper}$ with probability $1$ and so $S_{FPRD}(\rho_\text{improper}) = 0$ as well.

Models of finite precision readout devices whose effects have non-trivial probabilities even
for improper mixed states, with the readout for a given state having a probability distribution dominated by nearby states, could also be considered. These are more physically plausible if they arise from a physical theory such as semi-classical gravity.   
However, any finite precision readout device models should replicate infinite-precision readout devices in the limit as the precision becomes large, so we need not fix on any particular finite-precision readout device model here. For simplicity, we shall not consider these probabilistic finite precision readout devices in our discussion (although they could be incorporated by altering some of the definitions and statements below).

The information theory underlying the thermodynamic analysis of a model with readout devices depends, inter alia, on the resources accessible (in principle or in practice), on the precision and type(s) of readout device(s), and on the precision of quantum state preparations and measurements. 
In this work, we limit ourselves to the consideration of readout devices that are either infinite precision or can be made arbitrarily (finitely) precise.  We also limit ourselves to discussing finite ensembles. 
Under these assumptions, all quantum measurements can be represented (to arbitrary precision) 
as coarse-grained measurements using readout devices and  random number generators (by assumption arbitrarily precise).      

In a model with infinite precision readout devices, microsystems can be understood as being classical in the sense that every possible quantum state represents a different and perfectly distinguishable physical state.   If we consider a system of quantum states that is allowed to diffuse and is subject to readout device measurements (but not quantum measurements), it behaves like a classical ideal gas of chemical species.  Under these circumstances, a proper mixture of states behaves thermodynamically as a mixture of distinct chemical species, while an improper mixture behaves as a single species.   From this perspective, the measurement entropy \eqref{measurement entropy} is clearly a more natural measure than the decomposition entropy \eqref{decomposition entropy}.

However, the model also includes quantum measurements of quantum states.   These have the usual probabilistic properties, and may either increase or decrease the measurement entropy of an ensemble, depending on the ensemble and the measurement choice.
Although the possibility of entropy decrease may seem problematic, a full thermodynamic model would need to consider also the entropy of readout devices and quantum measurement devices.

We will not fix a particular model of quantum measurement in readout device world.
It is enough for our discussion below to note that, in readout device world, when a thermodynamic cycle includes a quantum measurement, one needs to consider contributions to the measurement entropy from outside the measured system, and that there are sensible models in which these contributions compensate for any loss of system measurement entropy. Moreover, as we will see in the case of von Neumann's engine, although steps of a thermodynamical cycle involving quantum measurements may reduce the measurement entropy (if one does not take into consideration the entropy flow to the quantum measurement device), to close the cycle there must be a step involving a reading of a RD that increases the measurement entropy at least as much, and so there is no violation of the second law.

Finite-precision readout device models can be understood either by considering microsystems as having ``smeared classicality" or by reducing the number of states in the state space, depending on the underlying model of finite precision.  Again, the significance of information in such models can vary depending on the other available resources. For example, if one works in a model for which the accessible quantum states are qubits, with access to readout devices that indicate, through a single reading, that the quantum state of the system is contained in a specific region of the Bloch ball, then if one can apply arbitrary unitaries and can repeat the readings of the readout device, one may eventually gather enough information by rotating the Bloch ball to determine the quantum state to arbitrary precision.

Both infinite and finite precision readout device models are idealisations of the type of behaviour that consistent nonlinear versions of quantum theory would imply.    In an actual nonlinear theory that allows some form of effective readout device we generally expect other contributions to the total measurement entropy coming from sources external to the system analysed.   These depend on the details of the theory, but in general would include contributions arising from the internal structure and workings of the effective readout devices.   For example, they should include the classical phase space uncertainty of the gravitational field configurations in a semiclassical gravity theory.
As noted above, a complete model would also need to describe the interplay between quantum operations and readout devices.

These issues preclude a general model-independent discussion of the thermodynamics of readout devices. 
However, we will not need to discuss all contributions to entropy flow in all possible models, since the redefinitions of entropy described above suffice to rebut the no-go arguments raised in the literature against nonlinear extensions to quantum theory, as we now explain.

\section{Peres' argument}

As Peres notes \cite{Peres1993}, at first sight it seems that versions of quantum theory with a nonlinear modification of the Schrödinger equation inevitably violate the second law. To see this, consider two pure states $\ket{\alpha(t)}$ and $\ket{\beta(t)}$. Let 
\begin{equation}
    \label{eqn:den op}
    \rho(t) = \epsilon \ket{\alpha(t)}\bra{\alpha(t)} + (1-\epsilon) \ket{\beta(t)}\bra{\beta(t)}
\end{equation}
be a mixed state with $0 < \epsilon < 1$ and non-vanishing eigenvalues
\begin{equation}
    \label{eqn:eigens}
    \lambda_{1,2}(t) = \frac{1}{2} \pm \sqrt{\frac{1}{4}-\epsilon(1-\epsilon)(1-x(t))}
\end{equation}
where $x(t) := \abs{\braket{\alpha(t)}{\beta(t)}}^2$. The von Neumann entropy of this mixture $S_{VN}(t) = - \sum_i \lambda_i(t) \log(\lambda_i(t))$ satisfies $\frac{dS_{VN}}{dx(t)}<0$ for all $\epsilon$ and $t$. Thus,
\begin{align}
    \forall t,  \frac{dS_{VN}}{dt}(t) = \frac{dS_{VN}}{dx(t)} \frac{dx(t)}{dt} &\geq 0 \\ \Leftrightarrow \forall t,  \frac{dx(t)}{dt} &\leq 0 \\ \Rightarrow \forall t \geq 0, \abs{\braket{\alpha(t)}{\beta(t)}}^2 &\leq \abs{\braket{\alpha(0)}{\beta(0)}}^2 \, . 
\end{align}
Hence if $\exists t \geq 0$ such that $\abs{\braket{\alpha(t)}{\beta(t)}}^2 > \abs{\braket{\alpha(0)}{\beta(0)}}^2$ then $\exists t \geq 0$ such that $\frac{dx(t)}{dt} > 0$  and  $\frac{dS_{VN}}{dt}(t) < 0$.
Consider a complete orthogonal set $\{\ket{\alpha_k(t)}\}$ where, for every $\ket{\beta(t)}$ \begin{equation}
    \sum_k \abs{\braket{\alpha_k(t)}{\beta(t)}}^2 = 1
\end{equation}
Thus, if there is some m for which $\abs{\braket{\alpha_m(t)}{\beta(t)}}^2 < \abs{\braket{\alpha_m(0)}{\beta(0)}}^2$ then there must also be some n for which $\abs{\braket{\alpha_n(t)}{\beta(t)}}^2 > \abs{\braket{\alpha_n(0)}{\beta(0)}}^2$, i.e. the entropy of $\rho_n(t) = \epsilon \ket{\alpha_n(t)}\bra{\alpha_n(t)} + (1-\epsilon) \ket{\beta(t)}\bra{\beta(t)}$ would spontaneously decrease and thus the second law is violated unless $\abs{\braket{\alpha_k(t)}{\beta(t)}}^2 = \abs{\braket{\alpha_k(0)}{\beta(0)}}^2$ for every $\ket{\alpha_k(t)}$ and $\ket{\beta(t)}$ and all $t$, which, from Wigner's theorem, implies that time evolution is unitary or anti-unitary, with the latter excluded by continuity. Thus, unless the evolution equation is linear, the second law is violated. 
Of course, in quantum theory, the Schr\"odinger evolution is linear, and orthogonal states remain orthogonal: i.e., $\braket{\alpha(0)}{\beta(0)} = 0 \Rightarrow  \braket{\alpha(t)}{\beta(t)} = 0$ for all $t$.  Hence this issue does not arise.   

However, this argument for second law violation in nonlinear theories assumes that other postulates that apply to quantum theory remain unchanged in such theories. In particular, it assumes that von Neumann entropy represents thermodynamical entropy.
This need not necessarily be the case in nonlinear modifications of quantum theory. It also assumes that the time evolution for density operators follows equation \eqref{eqn:den op}.
This need not be the case either for improper mixtures, as we now review.

\subsection{Nonlinear time evolution in readout device world and the Mixture Equivalence Principle}

There are two kinds of state mixing in quantum theory: improper mixing, which arises as a description of a subsystem obtaining from tracing out a degree of freedom of an entangled multipartite system, and proper mixing, which arises from (classical) statistical considerations of an ensemble of states. 
It is important to highlight that, in nonlinear extensions of quantum theory, proper and improper mixtures are generally distinguishable: the so-called mixture equivalence principle \cite{Fedida2025} fails to hold. 

In the case of proper mixtures, as was argued in \cite{Fedida2025}, a proper mixture $\rho(0)$ based on an ensemble $\{p_i,\ket{\psi_i(0)}\}$ indeed evolves to a proper mixture at time $t$ to $\rho(t)$ based on an  ensemble $\{p_i,\ket{\psi_i(t)}\}$, even if the time-evolution is nonlinear. That is, equation \eqref{eqn:den op} does hold if $\rho(0)$ is a proper mixture of $\ket{\alpha (0)}$ and $\ket{\beta (0)}$.

However, an initial improper mixture whose density matrix $\rho(0)$
can be written in the form $ \rho(0) = \sum_i p_i \ket{\psi_i(0)}\bra{\psi_i(0)}$ will \emph{not} generally evolve to $\sum_i p_i \ket{\psi_i(t)}\bra{\psi_i(t)}$.  
The time evolution of an improper mixture $\rho$ depends on that of the pure state $\ket{\psi} \in \mathcal{H}_1 \otimes \mathcal{H}_2$ of which it is the reduced density operator, with $\rho(t) = \Tr_{\mathcal{H}_2}(\ket{\psi(t)}\bra{\psi(t)})$. 
Theories with readout devices trivially allow nonlinearity \cite{Kent2005}, meaning $\rho(t)$ will generally not be of the form of equation \eqref{eqn:den op} for $t>0$ when $\rho$ is improper.

\subsection{Peres' argument, revisited}

Peres' argument relies on subtle assumptions that do not hold in nonlinear extensions of quantum theory. Foremost, we must now use measurement entropy instead of von Neumann entropy as the relevant entropic quantifier for a second law. In the case of proper mixtures, we can use the fact that measurement entropy in readout device world is always
\begin{equation}
    S_{RD}(\{p_i,\ket{\psi_i(t)}\}) = - \sum_i p_i \log(p_i)    
\end{equation}
and deduce that, in Peres' thought experiment, $S_{RD}$ is constant and equal to $- \epsilon \log(\epsilon) + (1-\epsilon) \log(1-\epsilon)$ for all $t \geq 0$. Importantly, note that in this case the measurement and decomposition entropy of $\rho$ are not equal, the latter being irrelevant for the discussion relevant to operationality and the second law. There is thus no violation of a second law of thermodynamics in the proper mixed state case.

In the case of improper mixtures, the analysis is different. As argued above, an improper mixture $\rho(0)$ does not generally evolve to the $\rho(t)$ defined by equation \eqref{eqn:den op} at later times. Nevertheless, regardless of whether one has access to an infinite-precision or a finite-precision RD, $S_{(FP)RD}(\rho_\text{improper}(t)) = 0$ for all times. There is thus no violation of a generalised second law in this case either.

\section{von Neumann's argument}

We now review the thought experiment originally discussed by von Neumann \cite{VonNeumann1955,Peres1993}. It takes the form of a thermodynamic cycle, shown in Figure \ref{fig:von Neumann Peres}. It relies on the existence of semipermeable walls: boundaries between two regions that allow some states to go through but not others.
In von Neumann's words \cite{VonNeumann1955}, for two quantum states $\ket{\phi},\ket{\psi} \in \mathcal{H}$, ``if $\ket{\phi},\ket{\psi}$ are not orthogonal then the assumed existence of such a wall would contradict the second law of thermodynamics". To quote Peres \cite{Peres1993}, we have become ``wily inventor[s] [who] claim having produced semipermeable partitions which unambiguously distinguish non-orthogonal states. [We] can thereby convert into work an unlimited amount of heat extracted from an isothermal reservoir [...] Will you invest your money in this new technology?" 

\begin{figure}[t!]
    \centering
    \begin{tikzpicture}[x=0.75pt,y=0.75pt,yscale=-1,xscale=1]

\draw   (239.33,21) -- (289.33,21) -- (289.33,71) -- (239.33,71) -- cycle ;
\draw   (320.67,20) -- (370.67,20) -- (370.67,70) -- (320.67,70) -- cycle ;
\draw   (381.51,211.71) -- (381.76,161.71) -- (431.76,161.96) -- (431.51,211.96) -- cycle ;
\draw   (380.91,130.38) -- (381.15,80.38) -- (431.15,80.62) -- (430.91,130.62) -- cycle ;
\draw    (406,134.8) -- (406.03,153.3) ;
\draw [shift={(406.03,155.3)}, rotate = 269.91] [color={rgb, 255:red, 0; green, 0; blue, 0 }  ][line width=0.75]    (7.65,-2.3) .. controls (4.86,-0.97) and (2.31,-0.21) .. (0,0) .. controls (2.31,0.21) and (4.86,0.98) .. (7.65,2.3)   ;
\draw    (382.67,46.13) -- (402.65,67.99) ;
\draw [shift={(404,69.47)}, rotate = 227.56] [color={rgb, 255:red, 0; green, 0; blue, 0 }  ][line width=0.75]    (7.65,-2.3) .. controls (4.86,-0.97) and (2.31,-0.21) .. (0,0) .. controls (2.31,0.21) and (4.86,0.98) .. (7.65,2.3)   ;
\draw    (316,248.13) -- (296.03,248.59) ;
\draw [shift={(294.03,248.63)}, rotate = 358.7] [color={rgb, 255:red, 0; green, 0; blue, 0 }  ][line width=0.75]    (7.65,-2.3) .. controls (4.86,-0.97) and (2.31,-0.21) .. (0,0) .. controls (2.31,0.21) and (4.86,0.98) .. (7.65,2.3)   ;
\draw    (294.67,45.47) -- (314,46.07) ;
\draw [shift={(316,46.13)}, rotate = 181.79] [color={rgb, 255:red, 0; green, 0; blue, 0 }  ][line width=0.75]    (7.65,-2.3) .. controls (4.86,-0.97) and (2.31,-0.21) .. (0,0) .. controls (2.31,0.21) and (4.86,0.98) .. (7.65,2.3)   ;
\draw    (229.33,248.8) -- (207.28,222.34) ;
\draw [shift={(206,220.8)}, rotate = 50.19] [color={rgb, 255:red, 0; green, 0; blue, 0 }  ][line width=0.75]    (7.65,-2.3) .. controls (4.86,-0.97) and (2.31,-0.21) .. (0,0) .. controls (2.31,0.21) and (4.86,0.98) .. (7.65,2.3)   ;
\draw    (204,156.13) -- (204,137.47) ;
\draw [shift={(204,135.47)}, rotate = 90] [color={rgb, 255:red, 0; green, 0; blue, 0 }  ][line width=0.75]    (7.65,-2.3) .. controls (4.86,-0.97) and (2.31,-0.21) .. (0,0) .. controls (2.31,0.21) and (4.86,0.98) .. (7.65,2.3)   ;
\draw    (205.33,70.13) -- (226.65,46.94) ;
\draw [shift={(228,45.47)}, rotate = 132.58] [color={rgb, 255:red, 0; green, 0; blue, 0 }  ][line width=0.75]    (7.65,-2.3) .. controls (4.86,-0.97) and (2.31,-0.21) .. (0,0) .. controls (2.31,0.21) and (4.86,0.98) .. (7.65,2.3)   ;
\draw  [color={rgb, 255:red, 126; green, 211; blue, 33 }  ,draw opacity=1 ][fill={rgb, 255:red, 126; green, 211; blue, 33 }  ,fill opacity=1 ] (249.67,21.67) -- (279.33,21.67) -- (279.33,70.13) -- (249.67,70.13) -- cycle ;
\draw    (264.33,21.67) -- (264.33,70.33) ;
\draw  [color={rgb, 255:red, 126; green, 211; blue, 33 }  ,draw opacity=1 ][fill={rgb, 255:red, 126; green, 211; blue, 33 }  ,fill opacity=1 ] (322,20.77) -- (370,20.77) -- (370,69.23) -- (322,69.23) -- cycle ;
\draw    (345.67,20.67) -- (345.67,69.33) ;
\draw  [color={rgb, 255:red, 126; green, 211; blue, 33 }  ,draw opacity=1 ][fill={rgb, 255:red, 126; green, 211; blue, 33 }  ,fill opacity=1 ] (382.03,81.26) -- (430.03,81.26) -- (430.03,129.73) -- (382.03,129.73) -- cycle ;
\draw  [dash pattern={on 4.5pt off 4.5pt}]  (407.53,81.16) -- (407.53,129.83)(404.53,81.16) -- (404.53,129.83) ;
\draw  [color={rgb, 255:red, 126; green, 211; blue, 33 }  ,draw opacity=1 ][fill={rgb, 255:red, 126; green, 211; blue, 33 }  ,fill opacity=1 ] (382.67,162.7) -- (402.67,162.7) -- (402.67,211.17) -- (382.67,211.17) -- cycle ;
\draw  [dash pattern={on 4.5pt off 4.5pt}]  (408.13,162.7) -- (408.13,211.37)(405.13,162.7) -- (405.13,211.37) ;
\draw   (321.51,270.38) -- (321.76,220.38) -- (371.76,220.62) -- (371.51,270.62) -- cycle ;
\draw  [color={rgb, 255:red, 126; green, 211; blue, 33 }  ,draw opacity=1 ][fill={rgb, 255:red, 126; green, 211; blue, 33 }  ,fill opacity=1 ] (322.67,221.37) -- (342.67,221.37) -- (342.67,269.84) -- (322.67,269.84) -- cycle ;
\draw  [dash pattern={on 4.5pt off 4.5pt}]  (348.13,221.37) -- (348.13,270.04)(345.13,221.37) -- (345.13,270.04) ;
\draw   (239.51,271.05) -- (239.76,221.05) -- (289.76,221.29) -- (289.51,271.29) -- cycle ;
\draw  [color={rgb, 255:red, 126; green, 211; blue, 33 }  ,draw opacity=1 ][fill={rgb, 255:red, 126; green, 211; blue, 33 }  ,fill opacity=1 ] (240.67,222.04) -- (260.67,222.04) -- (260.67,270.5) -- (240.67,270.5) -- cycle ;
\draw    (266.13,221.87) -- (266.13,248.17) -- (266.13,270.54)(263.13,221.87) -- (263.13,248.17) -- (263.13,270.54) ;
\draw  [color={rgb, 255:red, 126; green, 211; blue, 33 }  ,draw opacity=1 ][fill={rgb, 255:red, 126; green, 211; blue, 33 }  ,fill opacity=1 ] (180.03,159.93) -- (204.03,159.93) -- (204.03,208.4) -- (180.03,208.4) -- cycle ;
\draw   (178.67,80.33) -- (228.67,80.33) -- (228.67,130.33) -- (178.67,130.33) -- cycle ;
\draw  [color={rgb, 255:red, 126; green, 211; blue, 33 }  ,draw opacity=1 ][fill={rgb, 255:red, 126; green, 211; blue, 33 }  ,fill opacity=1 ] (182.67,81.33) -- (212.33,81.33) -- (212.33,129.8) -- (182.67,129.8) -- cycle ;
\draw    (203.67,81) -- (203.67,129.67) ;
\draw    (254.67,54.33) -- (254.67,40.47) ;
\draw [shift={(254.67,37.47)}, rotate = 90] [fill={rgb, 255:red, 0; green, 0; blue, 0 }  ][line width=0.08]  [draw opacity=0] (5.36,-2.57) -- (0,0) -- (5.36,2.57) -- cycle    ;
\draw [shift={(254.67,57.33)}, rotate = 270] [fill={rgb, 255:red, 0; green, 0; blue, 0 }  ][line width=0.08]  [draw opacity=0] (5.36,-2.57) -- (0,0) -- (5.36,2.57) -- cycle    ;
\draw    (269.14,54.27) -- (277.52,43.19) ;
\draw [shift={(279.33,40.8)}, rotate = 127.1] [fill={rgb, 255:red, 0; green, 0; blue, 0 }  ][line width=0.08]  [draw opacity=0] (5.36,-2.57) -- (0,0) -- (5.36,2.57) -- cycle    ;
\draw [shift={(267.33,56.67)}, rotate = 307.1] [fill={rgb, 255:red, 0; green, 0; blue, 0 }  ][line width=0.08]  [draw opacity=0] (5.36,-2.57) -- (0,0) -- (5.36,2.57) -- cycle    ;
\draw    (330.67,54.33) -- (330.67,40.47) ;
\draw [shift={(330.67,37.47)}, rotate = 90] [fill={rgb, 255:red, 0; green, 0; blue, 0 }  ][line width=0.08]  [draw opacity=0] (5.36,-2.57) -- (0,0) -- (5.36,2.57) -- cycle    ;
\draw [shift={(330.67,57.33)}, rotate = 270] [fill={rgb, 255:red, 0; green, 0; blue, 0 }  ][line width=0.08]  [draw opacity=0] (5.36,-2.57) -- (0,0) -- (5.36,2.57) -- cycle    ;
\draw    (354.48,53.61) -- (362.86,42.53) ;
\draw [shift={(364.67,40.13)}, rotate = 127.1] [fill={rgb, 255:red, 0; green, 0; blue, 0 }  ][line width=0.08]  [draw opacity=0] (5.36,-2.57) -- (0,0) -- (5.36,2.57) -- cycle    ;
\draw [shift={(352.67,56)}, rotate = 307.1] [fill={rgb, 255:red, 0; green, 0; blue, 0 }  ][line width=0.08]  [draw opacity=0] (5.36,-2.57) -- (0,0) -- (5.36,2.57) -- cycle    ;
\draw    (415.81,112.94) -- (424.19,101.86) ;
\draw [shift={(426,99.47)}, rotate = 127.1] [fill={rgb, 255:red, 0; green, 0; blue, 0 }  ][line width=0.08]  [draw opacity=0] (5.36,-2.57) -- (0,0) -- (5.36,2.57) -- cycle    ;
\draw [shift={(414,115.33)}, rotate = 307.1] [fill={rgb, 255:red, 0; green, 0; blue, 0 }  ][line width=0.08]  [draw opacity=0] (5.36,-2.57) -- (0,0) -- (5.36,2.57) -- cycle    ;
\draw    (392.67,114.33) -- (392.67,100.47) ;
\draw [shift={(392.67,97.47)}, rotate = 90] [fill={rgb, 255:red, 0; green, 0; blue, 0 }  ][line width=0.08]  [draw opacity=0] (5.36,-2.57) -- (0,0) -- (5.36,2.57) -- cycle    ;
\draw [shift={(392.67,117.33)}, rotate = 270] [fill={rgb, 255:red, 0; green, 0; blue, 0 }  ][line width=0.08]  [draw opacity=0] (5.36,-2.57) -- (0,0) -- (5.36,2.57) -- cycle    ;
\draw    (388,183) -- (388,169.13) ;
\draw [shift={(388,166.13)}, rotate = 90] [fill={rgb, 255:red, 0; green, 0; blue, 0 }  ][line width=0.08]  [draw opacity=0] (5.36,-2.57) -- (0,0) -- (5.36,2.57) -- cycle    ;
\draw [shift={(388,186)}, rotate = 270] [fill={rgb, 255:red, 0; green, 0; blue, 0 }  ][line width=0.08]  [draw opacity=0] (5.36,-2.57) -- (0,0) -- (5.36,2.57) -- cycle    ;
\draw    (387.81,202.27) -- (396.19,191.19) ;
\draw [shift={(398,188.8)}, rotate = 127.1] [fill={rgb, 255:red, 0; green, 0; blue, 0 }  ][line width=0.08]  [draw opacity=0] (5.36,-2.57) -- (0,0) -- (5.36,2.57) -- cycle    ;
\draw [shift={(386,204.67)}, rotate = 307.1] [fill={rgb, 255:red, 0; green, 0; blue, 0 }  ][line width=0.08]  [draw opacity=0] (5.36,-2.57) -- (0,0) -- (5.36,2.57) -- cycle    ;
\draw    (329.76,240.67) -- (334.9,227.59) ;
\draw [shift={(336,224.8)}, rotate = 111.45] [fill={rgb, 255:red, 0; green, 0; blue, 0 }  ][line width=0.08]  [draw opacity=0] (5.36,-2.57) -- (0,0) -- (5.36,2.57) -- cycle    ;
\draw [shift={(328.67,243.47)}, rotate = 291.45] [fill={rgb, 255:red, 0; green, 0; blue, 0 }  ][line width=0.08]  [draw opacity=0] (5.36,-2.57) -- (0,0) -- (5.36,2.57) -- cycle    ;
\draw    (326.7,249.32) -- (339.3,255.48) ;
\draw [shift={(342,256.8)}, rotate = 206.05] [fill={rgb, 255:red, 0; green, 0; blue, 0 }  ][line width=0.08]  [draw opacity=0] (5.36,-2.57) -- (0,0) -- (5.36,2.57) -- cycle    ;
\draw [shift={(324,248)}, rotate = 26.05] [fill={rgb, 255:red, 0; green, 0; blue, 0 }  ][line width=0.08]  [draw opacity=0] (5.36,-2.57) -- (0,0) -- (5.36,2.57) -- cycle    ;
\draw    (244.03,253.32) -- (256.64,259.48) ;
\draw [shift={(259.33,260.8)}, rotate = 206.05] [fill={rgb, 255:red, 0; green, 0; blue, 0 }  ][line width=0.08]  [draw opacity=0] (5.36,-2.57) -- (0,0) -- (5.36,2.57) -- cycle    ;
\draw [shift={(241.33,252)}, rotate = 26.05] [fill={rgb, 255:red, 0; green, 0; blue, 0 }  ][line width=0.08]  [draw opacity=0] (5.36,-2.57) -- (0,0) -- (5.36,2.57) -- cycle    ;
\draw    (247.76,242.01) -- (252.9,228.93) ;
\draw [shift={(254,226.13)}, rotate = 111.45] [fill={rgb, 255:red, 0; green, 0; blue, 0 }  ][line width=0.08]  [draw opacity=0] (5.36,-2.57) -- (0,0) -- (5.36,2.57) -- cycle    ;
\draw [shift={(246.67,244.8)}, rotate = 291.45] [fill={rgb, 255:red, 0; green, 0; blue, 0 }  ][line width=0.08]  [draw opacity=0] (5.36,-2.57) -- (0,0) -- (5.36,2.57) -- cycle    ;
\draw    (187.76,188.67) -- (192.9,175.59) ;
\draw [shift={(194,172.8)}, rotate = 111.45] [fill={rgb, 255:red, 0; green, 0; blue, 0 }  ][line width=0.08]  [draw opacity=0] (5.36,-2.57) -- (0,0) -- (5.36,2.57) -- cycle    ;
\draw [shift={(186.67,191.47)}, rotate = 291.45] [fill={rgb, 255:red, 0; green, 0; blue, 0 }  ][line width=0.08]  [draw opacity=0] (5.36,-2.57) -- (0,0) -- (5.36,2.57) -- cycle    ;
\draw    (206.53,106.55) -- (219.14,112.72) ;
\draw [shift={(221.83,114.03)}, rotate = 206.05] [fill={rgb, 255:red, 0; green, 0; blue, 0 }  ][line width=0.08]  [draw opacity=0] (5.36,-2.57) -- (0,0) -- (5.36,2.57) -- cycle    ;
\draw [shift={(203.83,105.23)}, rotate = 26.05] [fill={rgb, 255:red, 0; green, 0; blue, 0 }  ][line width=0.08]  [draw opacity=0] (5.36,-2.57) -- (0,0) -- (5.36,2.57) -- cycle    ;
\draw    (192.43,114.01) -- (197.57,100.93) ;
\draw [shift={(198.67,98.13)}, rotate = 111.45] [fill={rgb, 255:red, 0; green, 0; blue, 0 }  ][line width=0.08]  [draw opacity=0] (5.36,-2.57) -- (0,0) -- (5.36,2.57) -- cycle    ;
\draw [shift={(191.33,116.8)}, rotate = 291.45] [fill={rgb, 255:red, 0; green, 0; blue, 0 }  ][line width=0.08]  [draw opacity=0] (5.36,-2.57) -- (0,0) -- (5.36,2.57) -- cycle    ;
\draw    (406,222.13) -- (383.39,245.37) ;
\draw [shift={(382,246.8)}, rotate = 314.22] [color={rgb, 255:red, 0; green, 0; blue, 0 }  ][line width=0.75]    (7.65,-2.3) .. controls (4.86,-0.97) and (2.31,-0.21) .. (0,0) .. controls (2.31,0.21) and (4.86,0.98) .. (7.65,2.3)   ;
\draw  [color={rgb, 255:red, 126; green, 211; blue, 33 }  ,draw opacity=1 ][fill={rgb, 255:red, 126; green, 211; blue, 33 }  ,fill opacity=0.32 ] (206.03,159.18) -- (230.03,159.18) -- (230.03,207.64) -- (206.03,207.64) -- cycle ;
\draw    (211.73,181.33) -- (224.34,187.49) ;
\draw [shift={(227.03,188.81)}, rotate = 206.05] [fill={rgb, 255:red, 0; green, 0; blue, 0 }  ][line width=0.08]  [draw opacity=0] (5.36,-2.57) -- (0,0) -- (5.36,2.57) -- cycle    ;
\draw [shift={(209.03,180.01)}, rotate = 26.05] [fill={rgb, 255:red, 0; green, 0; blue, 0 }  ][line width=0.08]  [draw opacity=0] (5.36,-2.57) -- (0,0) -- (5.36,2.57) -- cycle    ;
\draw    (205.53,159.93) -- (205.53,186.23) -- (205.53,208.6)(202.53,159.93) -- (202.53,186.23) -- (202.53,208.6) ;
\draw   (180.03,208.4) -- (180.28,158.4) -- (230.28,158.64) -- (230.03,208.64) -- cycle ;

\draw (257.33,75.67) node [anchor=north west][inner sep=0.75pt]  [font=\scriptsize] [align=left] {(a)};
\draw (337.33,75) node [anchor=north west][inner sep=0.75pt]  [font=\scriptsize] [align=left] {(b)};
\draw (360.67,99.67) node [anchor=north west][inner sep=0.75pt]  [font=\scriptsize] [align=left] {(c)};
\draw (362,179) node [anchor=north west][inner sep=0.75pt]  [font=\scriptsize] [align=left] {(d)};
\draw (338.67,203.67) node [anchor=north west][inner sep=0.75pt]  [font=\scriptsize] [align=left] {(e)};
\draw (258,204.33) node [anchor=north west][inner sep=0.75pt]  [font=\scriptsize] [align=left] {(f)};
\draw (234.67,181.67) node [anchor=north west][inner sep=0.75pt]  [font=\scriptsize] [align=left] {(g)};
\draw (233.33,101.67) node [anchor=north west][inner sep=0.75pt]  [font=\scriptsize] [align=left] {(h)};

\end{tikzpicture}
    \caption{\justifying Cycle extracting heat from an isothermal reservoir and converting it into work using a semi-permeable non-quantum membrane that separates a specific pair of non-orthogonal states 
 as well as a semi-permeable quantum membrane that separates a specific pair of orthogonal states. The quantum membranes, which let through one of two orthogonal states, are depicted as solid double lines, while the semi-permeable membranes, which can be implemented by readout devices and let through one of two non-orthogonal states, are shown in dashed double lines. Walls, which do not let any states through, are depicted with single lines.}
    \label{fig:von Neumann Peres}
\end{figure}

Suppose that there are N photons in a box that has two chambers of equal volume. Half of the photons are prepared with vertical linear polarisation and occupy one chamber, the other half with linear polarisation at 45° from the vertical and occupy the other chamber, as seen in (a). These chambers undergo an isothermal expansion which doubles their volume (b), supplying a work $W^+ = N k_B T\log(2)$, where T is the temperature of the reservoir. The walls separating the photon gases are then replaced by semipermeable membranes (c) which can select non-orthogonal states (*). 

One of the membranes is transparent to vertically polarised photons and reflects those polarised at 45°; the other membrane does the opposite. We use a double frictionless piston as in (d) to obtain a mixture
\begin{equation}
    \label{Mixed polarisation}
    \rho = \frac{1}{2}\Big(\ket{\updownarrow}\bra{\updownarrow} + \ket{\frac{\pi}{4}}\bra{\frac{\pi}{4}}\Big) = \begin{pmatrix}
        \frac{3}{4} && \frac{1}{4} \\ \frac{1}{4} && \frac{1}{4}
    \end{pmatrix} \, .
\end{equation}
The eigenvalues of $\rho$ correspond to photons polarised at $\frac{\pi}{8}$ radians from the vertical and those orthogonal (e). We then replace the semipermeable membranes with quantum mechanical ones (f), which separate the two orthogonal polarisation states (g). We replace the quantum mechanical membranes by a wall and isothermally compress the gas (h) where both chambers
have the same pressure and the same total volume as those in (a). The isothermal compression requires an expenditure of work of $W^- = N k_B T S(\rho) < W^+$ (**). We finish the cycle by unitarily rotating the polarisation of the photons in each chamber so as to complete the cycle (a)\footnote{If $\ket{\psi}$ and $\ket{\phi}$ are two orthogonal states of a quantum system, then $U = \exp[\lambda(\ket{\psi} \bra{\phi} - \ket{\phi} \bra{\psi})]$, with $\lambda \in \mathbb{R}$, is a unitary which rotates the subspace spanned by $\ket{\psi}$ and $\ket{\phi}$: $U\ket{\psi} = \cos(\lambda) \ket{\psi} - \sin(\lambda) \ket{\phi}$ and $U \ket{\phi} = \sin(\lambda) \ket{\psi} + \cos(\lambda) \ket{\phi}$ \cite{Peres1993}.}.  Thus, we have net decrease of the von Neumann entropy over a full cycle. 

However, this argument for second law violation again assumes that other postulates of quantum theory remain unchanged, and that the information-theoretic properties of the theory follow that of quantum theory. We now show that, although readout device models indeed allow perfect discrimination of non-orthogonal states, the von Neumann cycle does not demonstrate problematic thermodynamical behaviour.

\subsection{Selection of non-orthogonal states in readout device world}

Readout devices allow for what von Neumann argues is the key impossible step in his engine that stops quantum theory from violating a second law, that is, the selection of non-orthogonal states. This is easy to see with e.g. an (FP)PRD: given a basis and two non-orthogonal states $\ket{\psi}$ and $\ket{\phi}$, one retrieves the classical description of both.

Provided the precision $l$ is sufficient to distinguish both, an agent (a Maxwell demon) can then determine which of the vertical linear polarisation or those at $45^{\circ}$ from the vertical are incoming towards a membrane, and let them through or not at will.

For example, in semiclassical gravity, step (*) can be implemented as follows:
\begin{enumerate}
    \item Convert the polarisation degrees of freedom of the photons into position degrees of freedom (say, with a beam splitter) - this is (up to reversible nonlinear corrections) a unitary process which does not generate any entropy or work;
    \item Read out the position states using readout devices.  This produces classical records of the states in the device outputs, with each readout corresponding to one of two distinct classical strings representing the two original polarisation states.   The proportions of these strings in the list of readouts are the same as those of the polarisation states, i.e., equal. In the case where one must distinguish between two polarisation states, this generates one bit of information, the corresponding work is $W^{RD} = N k_B T \log(2)$ by Landauer's principle.  
    \item Reconvert the position degrees of freedom into polarisation degrees of freedom.  
\end{enumerate}

This amounts to performing a Cavendish experiment, writing down the value of the gravitational field sourced by the superposition of position eigenstates, and deduce the resulting expression of the state.

\subsection{von Neumann's argument, revisited}

From these readings of an RD, we see already that 
\begin{equation}
    W^+ - (W^- + W^{RD}) \leq 0 \, ,
\end{equation}
i.e. the second law of thermodynamics is not violated should the rest of the analysis cycle be maintained. Thus, one could argue that, through a ``realistic" implementation of a Maxwell demon, one already evades this violation of the second law in a similar fashion to how one evades Maxwell's demon in classical thermodynamics and that in quantum thermodynamics in Szilard's engine \cite{Szilard1929}.

However, for the sake of argument, we can check whether the rest of von Neumann's analysis still holds in readout device world. In step (**), specifically in the assumption that (d) is physically equivalent to (e), it is also assumed that the entropy of $\rho$ is the Shannon entropy given by its eigendecomposition. Neither assumption holds in readout device world, where 
distinct ensembles that result in the same proper mixture are distinguishable, so that (d) and (e) are not physically equivalent.    The entropy of the ensemble $\{p_i,\ket{\psi_i}\}$ is the Shannon entropy of that ensemble, \emph{even if it is not the eigendecomposition of the associated mixture}.
We thus have that
\begin{equation}
    S_{RD}(\rho) = -\frac{1}{2} \log\big(\frac{1}{2}\big) -\frac{1}{2} \log\big(\frac{1}{2}\big) = \log(2) \, .
\end{equation}
That is, $W^- = N k_B T \log(2) = W^+$, i.e. even without taking into account the entropy production of the readings (*), we would have that $W^+ - W^- = 0$ i.e. no second law of thermodynamics is broken. This is true even if $\rho$ was an improper mixture: given a state readout device (whether finite precision or not), though the measurement entropy associated to the mixture would be 0, we would also have that $W^+ = W^- = 0$ and thus we still have from step $(*)$ that $W^+ - (W^- + W^{RD}) \leq 0$, so operationally no work is generated after one cycle. 

\subsection{Entropy reductions and the second law}

Because (d) and (e) are not equivalent in readout device world, the cycle should properly be understood as involving a step from (d) to (g).   To implement this requires quantum measurements to implement the required semi-permeable quantum membrane.   However, these measurements decrease the entropy of the system in readout device world. Indeed, if we start in (a) with a proper mixture, it follows that going from (d) to (g) - i.e. from a state which is ``more mixed" to one which is ``less mixed" - reduces the measurement entropy.

As discussed above, a complete model needs to consider the (readout device appropriate, i.e., measurement) entropy that is transferred to the quantum measurement device by the measurements. 
Without a complete description of the RD and quantum semi-permeable membranes and the entropy they transfer to the associated devices one cannot correctly calculate the entropy change over the cycle.

\begin{figure*}
    \centering
    \begin{tikzpicture}[x=0.75pt,y=0.75pt,yscale=-1,xscale=1]

\draw   (106.93,107) -- (209.33,107) -- (209.33,182.4) -- (106.93,182.4) -- cycle ;
\draw    (46.67,146.54) -- (108.3,146.3) ;
\draw [shift={(82.48,146.4)}, rotate = 179.78] [fill={rgb, 255:red, 0; green, 0; blue, 0 }  ][line width=0.08]  [draw opacity=0] (8.93,-4.29) -- (0,0) -- (8.93,4.29) -- cycle    ;
\draw    (209.73,118.21) -- (257.04,117.99) ;
\draw [shift={(260.04,117.97)}, rotate = 179.73] [fill={rgb, 255:red, 0; green, 0; blue, 0 }  ][line width=0.08]  [draw opacity=0] (8.93,-4.29) -- (0,0) -- (8.93,4.29) -- cycle    ;
\draw    (210.35,145.34) -- (257.67,145.11) ;
\draw [shift={(260.67,145.1)}, rotate = 179.73] [fill={rgb, 255:red, 0; green, 0; blue, 0 }  ][line width=0.08]  [draw opacity=0] (8.93,-4.29) -- (0,0) -- (8.93,4.29) -- cycle    ;
\draw    (210.04,171.86) -- (257.35,171.63) ;
\draw [shift={(260.35,171.62)}, rotate = 179.73] [fill={rgb, 255:red, 0; green, 0; blue, 0 }  ][line width=0.08]  [draw opacity=0] (8.93,-4.29) -- (0,0) -- (8.93,4.29) -- cycle    ;
\draw   (437.59,105.67) -- (588.35,105.67) -- (588.35,181.07) -- (437.59,181.07) -- cycle ;
\draw    (587.73,114.21) -- (635.04,113.99) ;
\draw [shift={(638.04,113.97)}, rotate = 179.73] [fill={rgb, 255:red, 0; green, 0; blue, 0 }  ][line width=0.08]  [draw opacity=0] (8.93,-4.29) -- (0,0) -- (8.93,4.29) -- cycle    ;
\draw    (588.35,141.34) -- (635.67,141.11) ;
\draw [shift={(638.67,141.1)}, rotate = 179.73] [fill={rgb, 255:red, 0; green, 0; blue, 0 }  ][line width=0.08]  [draw opacity=0] (8.93,-4.29) -- (0,0) -- (8.93,4.29) -- cycle    ;
\draw    (588.04,167.86) -- (635.35,167.63) ;
\draw [shift={(638.35,167.62)}, rotate = 179.73] [fill={rgb, 255:red, 0; green, 0; blue, 0 }  ][line width=0.08]  [draw opacity=0] (8.93,-4.29) -- (0,0) -- (8.93,4.29) -- cycle    ;
\draw   (462,146) -- (498,146) -- (498,175.07) -- (462,175.07) -- cycle ;
\draw    (498,168.4) -- (588.04,167.86) ;
\draw [shift={(548.02,168.1)}, rotate = 179.65] [fill={rgb, 255:red, 0; green, 0; blue, 0 }  ][line width=0.08]  [draw opacity=0] (8.93,-4.29) -- (0,0) -- (8.93,4.29) -- cycle    ;
\draw   (485.76,108) -- (513.33,108) -- (513.33,126.4) -- (485.76,126.4) -- cycle ;
\draw    (376,143.88) -- (437.63,143.63) ;
\draw [shift={(411.82,143.74)}, rotate = 179.78] [fill={rgb, 255:red, 0; green, 0; blue, 0 }  ][line width=0.08]  [draw opacity=0] (8.93,-4.29) -- (0,0) -- (8.93,4.29) -- cycle    ;
\draw   (531.36,122.67) -- (570.96,122.67) -- (570.96,152.8) -- (531.36,152.8) -- cycle ;
\draw    (437.63,143.63) -- (461.33,160.4) ;
\draw [shift={(453.56,154.9)}, rotate = 215.28] [fill={rgb, 255:red, 0; green, 0; blue, 0 }  ][line width=0.08]  [draw opacity=0] (8.93,-4.29) -- (0,0) -- (8.93,4.29) -- cycle    ;
\draw    (513.52,116.84) -- (530.16,125.2) ;
\draw [shift={(526.31,123.26)}, rotate = 206.68] [fill={rgb, 255:red, 0; green, 0; blue, 0 }  ][line width=0.08]  [draw opacity=0] (8.93,-4.29) -- (0,0) -- (8.93,4.29) -- cycle    ;
\draw    (571.6,144.93) -- (588.35,141.34) ;
\draw [shift={(584.87,142.09)}, rotate = 167.88] [fill={rgb, 255:red, 0; green, 0; blue, 0 }  ][line width=0.08]  [draw opacity=0] (8.93,-4.29) -- (0,0) -- (8.93,4.29) -- cycle    ;
\draw    (570.93,132.27) -- (587.73,114.21) ;
\draw [shift={(582.73,119.58)}, rotate = 132.93] [fill={rgb, 255:red, 0; green, 0; blue, 0 }  ][line width=0.08]  [draw opacity=0] (8.93,-4.29) -- (0,0) -- (8.93,4.29) -- cycle    ;
\draw    (498.27,157.6) -- (530.93,140.27) ;
\draw [shift={(519.02,146.59)}, rotate = 152.05] [fill={rgb, 255:red, 0; green, 0; blue, 0 }  ][line width=0.08]  [draw opacity=0] (8.93,-4.29) -- (0,0) -- (8.93,4.29) -- cycle    ;

\draw (114.22,116.22) node [anchor=north west][inner sep=0.75pt]  [font=\small] [align=left] {\begin{minipage}[lt]{58.83pt}\setlength\topsep{0pt}
\begin{center}
Quantum \\Measurement
\end{center}

\end{minipage}};
\draw (28.67,140.73) node [anchor=north west][inner sep=0.75pt]  [font=\scriptsize]  {$\rho _{\text{in}}$};
\draw (262,111.07) node [anchor=north west][inner sep=0.75pt]  [font=\scriptsize]  {$\rho _{\text{out}}$};
\draw (260,131) node [anchor=north west][inner sep=0.75pt]  [font=\scriptsize] [align=left] {classical\\outcome};
\draw (262.67,168.4) node [anchor=north west][inner sep=0.75pt]  [font=\scriptsize]  {$\text{input } ``\rho _{\text{in}} "$};
\draw (311.33,134.73) node [anchor=north west][inner sep=0.75pt]    {$\Longleftrightarrow $};
\draw (640.67,107.07) node [anchor=north west][inner sep=0.75pt]  [font=\scriptsize]  {$\rho_{\text{out}}$};
\draw (638,127) node [anchor=north west][inner sep=0.75pt]  [font=\scriptsize] [align=left] {classical\\outcome};
\draw (639.33,163.07) node [anchor=north west][inner sep=0.75pt]  [font=\scriptsize]  {$\text{input }``\rho_{\text{in}}"$};
\draw (469.33,154.33) node [anchor=north west][inner sep=0.75pt]  [font=\small] [align=left] {RD};
\draw (487.33,113) node [anchor=north west][inner sep=0.75pt]  [font=\scriptsize] [align=left] {RNG};
\draw (148,161.07) node [anchor=north west][inner sep=0.75pt]  [font=\scriptsize]  {$\mathcal{I}(\cdot)$};
\draw (354.67,140.73) node [anchor=north west][inner sep=0.75pt]  [font=\scriptsize]  {$\rho_{\text{in}}$};
\draw (541.47,130.07) node [anchor=north west][inner sep=0.75pt]  [font=\footnotesize] [align=left] {$\mathcal{I}(\cdot)$};

\end{tikzpicture}
\caption{\justifying Quantum measurements as a black box in RD world. A possible model of quantum measurements in this post-quantum setting is as a composite process of (i) an RD capturing the information of the input state and placing it in some register (perhaps within the measurement apparatus), (ii) the quantum instrument $\mathcal{I}(\cdot)$ associated to the quantum measurement, applying the non-selective quantum operation, and (iii) a random number generator (RNG), selecting one specific outcome at random.}
\label{fig:black box mmt}
\end{figure*}

However, even without a complete model of quantum measurements in RD world, we see that step (*) would compensate any decrease of the thermodynamic entropy that could arise from step (d) to (g) to complete the cycle, i.e. there is no violation of the second law over a cycle.   We expect that, in more general cycles, steps with quantum measurements can generate work, but 
completing any cycle requires some use of a RD that generates information about the quantum state and thus requires at least as much work.

One may be uneasy at the idea that decreases of the measurement entropy are possible in a fundamental theory -- even if, over a cycle, there is no decrease.  One view is that this is just a feature of RD theories, and that there is no possibility of overall work generation, so the second law is not violated.

A second view is that temporary information (understood in the context of measurement entropy) non-conservation is unaesthetic and unsatisfactory, albeit not immediately contradictory. 
On this view, one should give a precise (RD world model-dependent) model of quantum measurements 
that shows such decreases never happen when one considers the full setup, because a transfer of information from the system to the measurement device compensates for any apparent decrease.

For example, one could model quantum measurements as a composite black box process which includes an RD measurement device, for which there is always a register of the initial quantum state after the measurement that is inaccessible if one only has access to standard quantum resources. Such a setup is depicted in Fig. \ref{fig:black box mmt}.  The information about the initial quantum state is then always conserved, and so there is never any measurement entropy decrease upon quantum measurements.

Note that in this model information is still generated in (almost) trivial quantum measurements, for which the quantum instrument acts (approximately) as the identity on the quantum state. 
This implies a continuous production of information about the quantum state during continuous measurements, and indeed during free evolution if this is modelled as a continuous measurement process. This is arguably not completely unphysical, however. For example, in semiclassical gravity, during free evolution, the classical gravitational field still continuously reads the local quantum state, acting as a continuous readout device.  

Again, this discussion depends on the model and on the accessible physical resources.  
Also, finite precision models for both quantum measurements and readout devices may be more appropriate in some contexts.   Our point is that there are sensible models whereby even entropy reductions under quantum measurements can be averted.

In summary, on either view, von Neumann's engine is implementable in readout device world but does not demonstrate the violation of any thermodynamical law there.

\section{Hänggi and Wehner's argument}

We now review an argument of Hänggi and Wehner which shows that, under several assumptions reviewed below, a violation of the spin uncertainty principle implies a violation of the second law of thermodynamics \cite{Hanggi2013}. The ``impossible cycle" relies on the notion of \textit{entropic} or \textit{fine-grained} uncertainty relations \cite{Deutsch1983,Maassen1988,Oppenheim2010} which reformulate the standard Heisenberg uncertainty relations in a systematic fashion such that the bound does not depend on the state used for the computation of the expectation values of the standard uncertainty relations
\begin{equation}
    \label{Uncertainty principle}
    \Delta_\psi \hat{A} \Delta_\psi \hat{B} \geq \frac{1}{2} \abs{\expval{\comm{\hat{A}}{\hat{B}}}{\psi}} \, ,  
\end{equation}
where 
\begin{equation}
    \Delta_\psi \hat{O} = \sqrt{\expval{\hat{O}^2}{\psi}-\expval{\hat{O}}{\psi}^2} \, .
\end{equation}
Let $\mathbf{f}$ and $\mathbf{g}$ be two eigenbases for projective measurements with pure effects $\{f_0,f_1\}$ and $\{g_0,g_1\}$, respectively. For example, in the X and Z bases these are $\{\ket{+}\bra{+},\ket{-}\bra{-}\}$ and $\{\ket{0}\bra{0},\ket{1}\bra{1}\}$ respectively. Fine-grained uncertainty relations for such measurements are then defined to be the following four inequalities
\begin{align}
    \label{Fine grained 00}
   \frac{1}{2}\Big(p(f_0|\ket{\psi}) + p(g_0|\ket{\psi})\Big) &\leq \zeta_{f_0,g_0} \, , \\
   \label{Fine grained 01}
   \frac{1}{2}\Big(p(f_0|\ket{\psi}) + p(g_1|\ket{\psi})\Big) &\leq \zeta_{f_0,g_1} \, , \\
   \label{Fine grained 10}
   \frac{1}{2}\Big(p(f_1|\ket{\psi}) + p(g_0|\ket{\psi})\Big) &\leq \zeta_{f_1,g_0} \, , \\
   \label{Fine grained 11}
   \frac{1}{2}\Big(p(f_1|\ket{\psi}) + p(g_1|\ket{\psi})\Big) &\leq \zeta_{f_1,g_1} \, ,
\end{align}
for any pure quantum state $\ket{\psi}$, where for a projective measurement in the Z basis with outcome $\ket{0}$, $p(\ket{0}|\ket{\psi}) = \abs{\braket{0}{\psi}}^2$ by the Born rule, and $\zeta_{f_i,g_j}>0$ are suprema over states for the measurement outcomes $\{f_i,g_j\}$. For $\zeta < 1$ these inequalities quantify uncertainty since if the outcome is certain for one outcome, then it cannot be for the other. For quantum theory which respects the Heisenberg uncertainty principle, the constraint on $\zeta$ is stronger, with $\zeta \leq \frac{1}{2} + \frac{1}{2\sqrt{2}}$. 

\subsection{Readout devices and the uncertainty principle}

First, we highlight that readout devices indeed violate the uncertainty principle. It is easy to see that given any two states, the associated fine-grained uncertainty relations are upper-bounded by $1$. For example, given $\rho_{0_X} = \ket{+}\bra{+}$ and $\rho_{1_Z} = \ket{1}\bra{1}$, there exists an effect $e$ such that
\begin{equation}
    \frac{1}{2} + \frac{1}{2\sqrt{2}} \leq \frac{1}{2}(p(e| \rho_{0_X}) + p(e|\rho_{1_Z})) = 1
\end{equation}
which can be achieved e.g. by an RD, or by letting the states evolve nonlinearly to orthogonal states and measuring quantum mechanically in an appropriate basis, which allows one to get arbitrarily close to the upper bound of $1$. 
Note that in readout device models, unlike the kind of theory considered by Hänggi and Wehner, the converse uncertainty relations defined by fixing a state and considering ``complementary" measurements (e.g. through projectors $\ket{+}\bra{+}$ and $\ket{0}\bra{0}$) cannot generally be formulated. Indeed, in readout device world, pure effects are not generally dual to pure states.

\subsection{The cycle}

Let us outline the steps of this engine using quantum notation. The cycle is presented in Figure \ref{fig:Hanggi-Wehner}.

\begin{figure*}
   \centering
   \begin{tikzpicture}[x=0.75pt,y=0.75pt,yscale=-0.95,xscale=0.95]
\draw   (45.33,40.33) -- (198.33,40.33) -- (198.33,115.33) -- (45.33,115.33) -- cycle ;
\draw   (258.67,40.33) -- (411.67,40.33) -- (411.67,115.33) -- (258.67,115.33) -- cycle ;
\draw   (468.67,41) -- (621.67,41) -- (621.67,116) -- (468.67,116) -- cycle ;
\draw   (258.67,176.33) -- (411.67,176.33) -- (411.67,251.33) -- (258.67,251.33) -- cycle ;
\draw   (469.33,175.67) -- (622.33,175.67) -- (622.33,250.67) -- (469.33,250.67) -- cycle ;
\draw    (210.67,80) -- (245.67,80) ;
\draw [shift={(247.67,80)}, rotate = 180] [color={rgb, 255:red, 0; green, 0; blue, 0 }  ][line width=0.75]    (10.93,-3.29) .. controls (6.95,-1.4) and (3.31,-0.3) .. (0,0) .. controls (3.31,0.3) and (6.95,1.4) .. (10.93,3.29)   ;
\draw    (423.33,80.67) -- (458.33,80.67) ;
\draw [shift={(460.33,80.67)}, rotate = 180] [color={rgb, 255:red, 0; green, 0; blue, 0 }  ][line width=0.75]    (10.93,-3.29) .. controls (6.95,-1.4) and (3.31,-0.3) .. (0,0) .. controls (3.31,0.3) and (6.95,1.4) .. (10.93,3.29)   ;
\draw    (248,213.33) -- (214.33,213.33) ;
\draw [shift={(212.33,213.33)}, rotate = 360] [color={rgb, 255:red, 0; green, 0; blue, 0 }  ][line width=0.75]    (10.93,-3.29) .. controls (6.95,-1.4) and (3.31,-0.3) .. (0,0) .. controls (3.31,0.3) and (6.95,1.4) .. (10.93,3.29)   ;
\draw    (462,213.33) -- (428.33,213.33) ;
\draw [shift={(426.33,213.33)}, rotate = 360] [color={rgb, 255:red, 0; green, 0; blue, 0 }  ][line width=0.75]    (10.93,-3.29) .. controls (6.95,-1.4) and (3.31,-0.3) .. (0,0) .. controls (3.31,0.3) and (6.95,1.4) .. (10.93,3.29)   ;
\draw    (123.67,162.67) -- (123.67,129.33) ;
\draw [shift={(123.67,127.33)}, rotate = 90] [color={rgb, 255:red, 0; green, 0; blue, 0 }  ][line width=0.75]    (10.93,-3.29) .. controls (6.95,-1.4) and (3.31,-0.3) .. (0,0) .. controls (3.31,0.3) and (6.95,1.4) .. (10.93,3.29)   ;
\draw    (544.33,127.33) -- (544.33,158.67) ;
\draw [shift={(544.33,160.67)}, rotate = 270] [color={rgb, 255:red, 0; green, 0; blue, 0 }  ][line width=0.75]    (10.93,-3.29) .. controls (6.95,-1.4) and (3.31,-0.3) .. (0,0) .. controls (3.31,0.3) and (6.95,1.4) .. (10.93,3.29)   ;
\draw    (105.67,40) -- (105.67,114.67) ;
\draw  [dash pattern={on 4.5pt off 4.5pt}]  (300.33,41.33) -- (300.33,116) ;
\draw  [dash pattern={on 4.5pt off 4.5pt}]  (326.33,40.67) -- (326.33,80.67) -- (326.33,115.33) ;
\draw  [dash pattern={on 4.5pt off 4.5pt}]  (509.67,41.33) -- (509.67,116) ;
\draw  [dash pattern={on 4.5pt off 4.5pt}]  (553.67,40.67) -- (553.67,115.33) ;
\draw  [dash pattern={on 0.84pt off 2.51pt}]  (497,175.33) -- (497,250) ;
\draw  [dash pattern={on 0.84pt off 2.51pt}]  (591.67,175.33) -- (591.67,250) ;
\draw    (285.67,176) -- (285.67,250.67) ;
\draw    (321.67,176.67) -- (321.67,251.33) ;
\draw    (363,176.67) -- (363,230) -- (363,251.33) ;
\draw   (45.17,174.33) -- (198.17,174.33) -- (198.17,249.33) -- (45.17,249.33) -- cycle ;
\draw    (70.17,174) -- (70.17,248.67) ;
\draw    (95.17,174.67) -- (95.17,249.33) ;
\draw    (149,174.67) -- (149,228) -- (149,249.33) ;
\draw    (300.37,80.17) -- (282.7,80.63)(300.29,77.17) -- (282.63,77.63) ;
\draw [shift={(274.67,79.33)}, rotate = 358.51] [color={rgb, 255:red, 0; green, 0; blue, 0 }  ][line width=0.75]    (10.93,-3.29) .. controls (6.95,-1.4) and (3.31,-0.3) .. (0,0) .. controls (3.31,0.3) and (6.95,1.4) .. (10.93,3.29)   ;
\draw    (326.29,79.17) -- (342.96,78.72)(326.37,82.17) -- (343.04,81.72) ;
\draw [shift={(351,80)}, rotate = 178.45] [color={rgb, 255:red, 0; green, 0; blue, 0 }  ][line width=0.75]    (10.93,-3.29) .. controls (6.95,-1.4) and (3.31,-0.3) .. (0,0) .. controls (3.31,0.3) and (6.95,1.4) .. (10.93,3.29)   ;
\draw    (288.67,50.67) -- (309,50.06) ;
\draw [shift={(311,50)}, rotate = 178.29] [color={rgb, 255:red, 0; green, 0; blue, 0 }  ][line width=0.75]    (10.93,-3.29) .. controls (6.95,-1.4) and (3.31,-0.3) .. (0,0) .. controls (3.31,0.3) and (6.95,1.4) .. (10.93,3.29)   ;
\draw    (284.33,95.33) -- (299,104) -- (284.9,108.7) ;
\draw [shift={(283,109.33)}, rotate = 341.57] [color={rgb, 255:red, 0; green, 0; blue, 0 }  ][line width=0.75]    (10.93,-3.29) .. controls (6.95,-1.4) and (3.31,-0.3) .. (0,0) .. controls (3.31,0.3) and (6.95,1.4) .. (10.93,3.29)   ;
\draw    (341.76,47.98) -- (327.5,54.25) -- (341.22,61.33) ;
\draw [shift={(343,62.25)}, rotate = 207.3] [color={rgb, 255:red, 0; green, 0; blue, 0 }  ][line width=0.75]    (10.93,-3.29) .. controls (6.95,-1.4) and (3.31,-0.3) .. (0,0) .. controls (3.31,0.3) and (6.95,1.4) .. (10.93,3.29)   ;
\draw    (337.67,98.46) -- (317,97.62) ;
\draw [shift={(315,97.54)}, rotate = 2.32] [color={rgb, 255:red, 0; green, 0; blue, 0 }  ][line width=0.75]    (10.93,-3.29) .. controls (6.95,-1.4) and (3.31,-0.3) .. (0,0) .. controls (3.31,0.3) and (6.95,1.4) .. (10.93,3.29)   ;
\draw    (511.26,206.39) -- (497,212.67) -- (510.72,219.75) ;
\draw [shift={(512.5,220.67)}, rotate = 207.3] [color={rgb, 255:red, 0; green, 0; blue, 0 }  ][line width=0.75]    (10.93,-3.29) .. controls (6.95,-1.4) and (3.31,-0.3) .. (0,0) .. controls (3.31,0.3) and (6.95,1.4) .. (10.93,3.29)   ;
\draw    (577,204) -- (591.67,212.67) -- (577.56,217.37) ;
\draw [shift={(575.67,218)}, rotate = 341.57] [color={rgb, 255:red, 0; green, 0; blue, 0 }  ][line width=0.75]    (10.93,-3.29) .. controls (6.95,-1.4) and (3.31,-0.3) .. (0,0) .. controls (3.31,0.3) and (6.95,1.4) .. (10.93,3.29)   ;
\draw    (106.17,174.5) -- (106.17,249.17) ;
\draw    (57,207) -- (57,220.25) ;
\draw [shift={(57,222.25)}, rotate = 270] [color={rgb, 255:red, 0; green, 0; blue, 0 }  ][line width=0.75]    (6.56,-1.97) .. controls (4.17,-0.84) and (1.99,-0.18) .. (0,0) .. controls (1.99,0.18) and (4.17,0.84) .. (6.56,1.97)   ;
\draw    (82.5,207) -- (82.5,220.25) ;
\draw [shift={(82.5,222.25)}, rotate = 270] [color={rgb, 255:red, 0; green, 0; blue, 0 }  ][line width=0.75]    (6.56,-1.97) .. controls (4.17,-0.84) and (1.99,-0.18) .. (0,0) .. controls (1.99,0.18) and (4.17,0.84) .. (6.56,1.97)   ;
\draw    (182,207.5) -- (182,220.75) ;
\draw [shift={(182,222.75)}, rotate = 270] [color={rgb, 255:red, 0; green, 0; blue, 0 }  ][line width=0.75]    (6.56,-1.97) .. controls (4.17,-0.84) and (1.99,-0.18) .. (0,0) .. controls (1.99,0.18) and (4.17,0.84) .. (6.56,1.97)   ;
\draw    (166.5,174.17) -- (166.5,227.5) -- (166.5,248.83) ;
\draw (107.17,16.33) node [anchor=north west][inner sep=0.75pt]   [align=left] {(a)};
\draw (320.67,16.17) node [anchor=north west][inner sep=0.75pt]   [align=left] {(b)};
\draw (526.17,16) node [anchor=north west][inner sep=0.75pt]   [align=left] {(c)};
\draw (536.67,254) node [anchor=north west][inner sep=0.75pt]   [align=left] {(d)};
\draw (324,252.67) node [anchor=north west][inner sep=0.75pt]   [align=left] {(e)};
\draw (108.17,252.17) node [anchor=north west][inner sep=0.75pt]   [align=left] {(f)};
\draw (62.33,86.07) node [anchor=north west][inner sep=0.75pt]  [font=\small]  {$\rho _{0}$};
\draw (150.5,86.4) node [anchor=north west][inner sep=0.75pt]  [font=\small]  {$\rho _{1}$};
\draw (51.33,46.4) node [anchor=north west][inner sep=0.75pt]  [font=\small]  {$V_{0}$};
\draw (170,45.07) node [anchor=north west][inner sep=0.75pt]  [font=\small]  {$V_{1}$};
\draw (288.76,117.56) node [anchor=north west][inner sep=0.75pt]  [font=\footnotesize]  {$M_{0}$};
\draw (318,117.23) node [anchor=north west][inner sep=0.75pt]  [font=\footnotesize]  {$M_{1}$};
\draw (268.17,47.73) node [anchor=north west][inner sep=0.75pt]  [font=\tiny]  {$\ket{e_{1}}$};
\draw (265,97.4) node [anchor=north west][inner sep=0.75pt]  [font=\tiny]  {$\ket{e_{0}}$};
\draw (346.17,49.23) node [anchor=north west][inner sep=0.75pt]  [font=\tiny]  {$\ket{e_{1}}$};
\draw (339,94.4) node [anchor=north west][inner sep=0.75pt]  [font=\tiny]  {$\ket{e_{0}}$};
\draw (499.43,117.83) node [anchor=north west][inner sep=0.75pt]  [font=\footnotesize]  {$M_{0}$};
\draw (548.67,117.57) node [anchor=north west][inner sep=0.75pt]  [font=\footnotesize]  {$M_{1}$};
\draw (481.33,70.9) node [anchor=north west][inner sep=0.75pt]  [font=\small]  {$\rho $};
\draw (524.83,71.23) node [anchor=north west][inner sep=0.75pt]  [font=\small]  {$\rho $};
\draw (581.33,71.23) node [anchor=north west][inner sep=0.75pt]  [font=\small]  {$\rho $};
\draw (486.43,156.33) node [anchor=north west][inner sep=0.75pt]  [font=\footnotesize]  {$M_{\phi _{j}}$};
\draw (513.76,214.79) node [anchor=north west][inner sep=0.75pt]  [font=\tiny]  {$\ket{\phi _{j}}$};
\draw (577.43,155.33) node [anchor=north west][inner sep=0.75pt]  [font=\footnotesize]  {$M_{\phi _{j}}$};
\draw (556.5,214.9) node [anchor=north west][inner sep=0.75pt]  [font=\tiny]  {$\ket{\phi _{j}}$};
\draw (261.5,208.4) node [anchor=north west][inner sep=0.75pt]  [font=\tiny]  {$\ket{\phi _{0}}$};
\draw (295.5,208.4) node [anchor=north west][inner sep=0.75pt]  [font=\tiny]  {$\ket{\phi _{1}}$};
\draw (377,208.9) node [anchor=north west][inner sep=0.75pt]  [font=\tiny]  {$\ket{\phi _{k}}$};
\draw (332.5,199) node [anchor=north west][inner sep=0.75pt]   [align=left] {...};
\draw (48,196.9) node [anchor=north west][inner sep=0.75pt]  [font=\tiny]  {$\ket{\phi _{0}}$};
\draw (72.5,196.4) node [anchor=north west][inner sep=0.75pt]  [font=\tiny]  {$\ket{\phi _{1}}$};
\draw (93,200) node [anchor=north west][inner sep=0.75pt]   [align=left] {...};
\draw (173.5,197.4) node [anchor=north west][inner sep=0.75pt]  [font=\tiny]  {$\ket{\phi _{k}}$};
\draw (119,200) node [anchor=north west][inner sep=0.75pt]   [align=left] {...};
\draw (50,227.9) node [anchor=north west][inner sep=0.75pt]  [font=\tiny]  {$\ket{\tau _{0}^{0}}$};
\draw (73.5,227.4) node [anchor=north west][inner sep=0.75pt]  [font=\tiny]  {$\ket{\tau _{0}^{0}}$};
\draw (174,227.4) node [anchor=north west][inner sep=0.75pt]  [font=\tiny]  {$\ket{\tau _{k}^{1}}$};
\draw (150,199.5) node [anchor=north west][inner sep=0.75pt]   [align=left] {...};
\end{tikzpicture}
   \caption{\justifying The Hänggi-Wehner cycle which can extract net work if the uncertainty principle is violated and other principles are left unchanged. As discussed in the text, this cycle fails to extract net work for the nonlinear models we consider. First (a) the system is prepared in two states $\rho_1$ and $\rho_2$ in separated volumes, each with $N/2$ particles. (b) We then replace the wall by semi-transparent membranes that let through one state but block the other. (c) These membranes move apart until equilibrium $\rho$ is reached. (d) Insert new membranes to separate the pure components of $\rho$. (e) Subdivide these into smaller regions such that the resulting pure states are building blocks of $\rho_1$ and $\rho_2$. (f) Unitarily transform these states into the pure state decomposition of $\rho_1$ and $\rho_2$}
   \label{fig:Hanggi-Wehner}
\end{figure*}

We start the engine with a box which contains two types of particles : $\rho_0$ and $\rho_1$, given by
\begin{align}
   \rho_0 &= \frac{1}{2} (\ket{+}\bra{+} + \ket{0}\bra{0}) = \frac{\mathbb{1} + \frac{X+Z}{2}}{2} \, , \\
   \rho_1 &= \frac{1}{2}(\ket{-}\bra{-} + \ket{1}\bra{1}) = \frac{\mathbb{1} - \frac{X+Z}{2}}{2} 
\end{align}
which are in two separated volumes. 
\begin{enumerate}
    \item[(a)] There are $N/2$ particles in the $\rho_0$ state and $N/2$ particles in the $\rho_1$ state (a).
    \item[(b)] We then make a projective measurement in the $\{\ket{e_0},\ket{e_1}\}$ basis where $\ket{e_0}$ and $\ket{e_1}$ are the eigenstates of $\frac{X+Z}{\sqrt{2}}$ (the measurement thus only has two outcomes), and replace the wall separating these two volumes by semi-transparent membranes $M_0$ and $M_1$ which let through $\ket{e_0}$ and $\ket{e_1}$ respectively, but block the other state.
    \item[(c)] These membranes then move apart until they reach equilibrium : we have the maximally mixed state
\begin{equation}
   \rho = \frac{1}{2}(\rho_0+\rho_1) = \frac{\mathbb{1}}{2}
\end{equation}
in the three regions. The work that can be extracted from this process is
\begin{equation}
\begin{split}
    W^+_1 = N k_B T \log(2) \Big[1 &- \frac{1}{2}H(\zeta_{\ket{+},\ket{0}}) \\ &-\frac{1}{2}H(\zeta_{\ket{-},\ket{1}})\Big] \, .
\end{split}    
\end{equation}
\item[(d)] We then insert new membranes on the side so as to separate the pure components of $\rho$ into regions which are only populated by pure states $\ket{\phi_j}$. This process requires a work
\begin{equation}
    W^- = Nk_B T \log(2) S(\rho) = N k_B T \log(2)
\end{equation}
as $S(\rho)=1$.
    \item[(e)] We then subdivide the volumes containing the pure states $\ket{\phi_j}$ into smaller volumes (e) such that the number of particles contained in these smaller volumes are proportional to $\frac{r_j^0}{2}$ and $\frac{r_j^1}{2}$, where
\begin{align}
    \rho_0 &= \sum_j r_j^0 \ket{\tau_j^0} \bra{\tau_j^0} \\
    \rho_1 &= \sum_j r_j^1 \ket{\tau_j^1} \bra{\tau_j^1} \, .
\end{align}
\item[(f)] The pure state $\ket{\phi_j}$ contained in each small volume is then unitarily transformed into the pure state $\ket{\tau_j^0}$ or $\ket{\tau_j^1}$, which are also pure and thus do not require work. We finally mix the different components of $\rho_0$ together (and likewise for $\rho_1$) so as to complete the cycle (a), which extracts work
\begin{equation}
    W^+_2 = N k_B T \log(2) \sum_{i=0}^1 \frac{1}{2} S(\rho_i) \, .
\end{equation}
Since $\rho_0$ and $\rho_1$ both have one positive eigenvalue $\frac{1}{2}+\frac{1}{2\sqrt{2}}$, we get
\begin{align}
    S(\rho_{0,1}) &= H\Big(\frac{1}{2}+\frac{1}{2\sqrt{2}}\Big) \\ &\equiv - \Big(\frac{1}{2}+\frac{1}{2\sqrt{2}}\Big) \log(\frac{1}{2}+\frac{1}{2\sqrt{2}}) \, .
\end{align}
Thus, the total work which can be extracted is
\begin{align}
    \Delta W &= W^+_1 + W^+_2 - W^- \\
    &= N k_B T \log(2) \Big[H\Big(\frac{1}{2} + \frac{1}{2\sqrt{2}}\Big) \nonumber \\ &-\frac{1}{2}H(\zeta_{\ket{+},\ket{0}}) - \frac{1}{2} H(\zeta_{\ket{-},\ket{1}}) \Big] \, .
\end{align}
\end{enumerate}

It is here evident that quantum theory does not violate the second law for such a cycle, but that any theory with quantum-like features that satisfies certain assumptions, for which $\zeta > \frac{1}{2} + \frac{1}{2\sqrt{2}}$, does violate the second law. However, again, we shall show that these assumptions are violated in the extensions of quantum theory that we consider here.

\subsection{Readout devices and Hänggi and Wehner's argument}

As was highlighted by Hänggi and Wehner in their original formulation of the engine \cite{Hanggi2013}, this cycle is based on several assumptions that the theory should respect:
\begin{enumerate}
    \item[A1] The state space is convex.
    \item[A2] Effects are linear functionals.
    \item[A3] Pure states are dual to pure effects. Uncertainty relations can thus be stated equivalently in terms of states or measurements\footnote{It was noted in \cite{Hanggi2013} that this assumption is satisfied by any theory which satisfies so-called bit-symmetry \cite{Muller2012}: every logical bit can be mapped to any other logical bit by a reversible transformation.}.
    \item[A4] Pure effects are projective.
    \item[A5] If $f_0 + f_1 = u$ for two effects $f_0$ and $f_1$, then the dual states $\rho_{f_0} + \rho_{f_1} = \rho_u$ and $e(\rho_u / 2) = \frac{1}{2}$ for all pure effects $e$.
    \item[A6] Let $\rho = \sum_{j=1}^d q_j \sigma_j$ be a decomposition of $\rho$ into perfectly distinguishable pure states $\sigma_j$. Let $h_{\sigma_j}$ denote the pure effect dual to $\sigma_j$. Then $\sum_{j=1}^d h_{\sigma_j} = u$ and $h_k(\sigma_j) = \delta_{kj}$ for all j and k.
    \item[A7] Let $\rho$ and $\sigma$ be pure states. Then the transformation $\rho \to \sigma$ is reversible (and thus does not require any work, neither can any work be gained from it).
\end{enumerate}

In readout device world, most of these assumptions, which are necessary for the derivation of the engine, are violated. Namely,
\begin{itemize}
    \item Assumption A2 is violated: readout devices output a classical description of density operators with probability $1$, i.e. they do not respect the convex structure of the state (while the Born rule does). 
    \item Assumptions A3, A5 and A6 are violated: pure states are not necessarily dual to pure effects because effects from readout devices are all pure. Consequently, A5 and A6 also fail to hold.
    \item Assumption A7 is violated: since we work with nonlinear quantum mechanics, it need not be the case that if $\rho$ and $\sigma$ are any pure states, the transformation $\rho \to \sigma$ is reversible. Hence, it may be that the transformation between two pure states requires work, or that work can be extracted from it.
\end{itemize}

Further note that the failure of A6 to hold also highlights the failure of the equivalence between decomposition entropy and measurement entropy, which are key in this derivation. Again, if one starts with a proper mixture given by an ensemble $\{1/2,\rho_0;1/2,\rho_1\}$, the entropy will be constantly that of the Shannon entropy of that specific ensemble. The subdivision of $\rho$ into different pure states does not induce a new value for the entropy because decomposition entropy is not the relevant quantity here. Likewise, if $\rho$ is an improper mixture, then the measurement entropy ought to be $0$ regardless of whether one works with infinite-precision or finite-precision RDs across the whole process unless one injects work into it. Thus, there is no violation of a generalised second law in this context.

\section{Conclusion}

We have seen that claims that nonlinear extensions of quantum theory necessarily violate the second law of thermodynamics need not hold once a generalisation to the measurement entropy, suitable for the theory, is considered. Focusing on a class of radical modifications, namely state readout devices, which can be realised for example by M{\o}ller-Rosenfeld semiclassical gravity, we demonstrated that several no-go arguments rest on implicit assumptions tied to standard quantum information theory. These include the identification of thermodynamic entropy with von Neumann entropy, the equivalence between proper and improper mixtures, and some form of duality between states and effects.

We re-evaluated classic arguments by Peres and von Neumann, showing that their conclusions do not hold in the context of readout device theories. In particular, von Neumann’s thought experiment, which purportedly extracts net work by separating non-orthogonal states, fails not only once one accounts for the work cost of information erasure via readout devices as required by Landauer's principle (which is a similar resolution to ``standard" Maxwellian and Szilardian demons), but also does not account for a more operational notion of entropy relevant to the nonlinear theory. Similarly, we addressed the Hänggi-Wehner engine, which links violations of the uncertainty principle to thermodynamic violations. We showed that the derivation of this result depends on several assumptions -- such as the linearity of effects and the equivalence between pure states and effects -- which are violated in readout device models. As such, violations of entropic uncertainty relations in these models need not lead to thermodynamic paradoxes.

These results suggest that nonlinear extensions of quantum theory -- when carefully constructed and interpreted -- can remain thermodynamically consistent, though we do not provide a model-independent proof. In particular, semiclassical gravity, which admits non-standard readout mechanisms without necessarily violating relativistic causality, provides a framework in which non-standard physical phenomena can arise without (at least immediately) contradicting thermodynamic principles. Thermodynamic arguments to date thus give no reason to reject such models.

\section{Acknowledgments}

We acknowledge financial support from the
UK Quantum Communications Hub grant no. 
EP/T001011/1  and project OPP640, funded by the Science and Technology Facilities Council's International Science Partnerships Fund.
S.F. is funded by a studentship from the Engineering and Physical Sciences Research Council. A.K. was supported in part by Perimeter Institute for
Theoretical Physics. Research at Perimeter Institute is supported by
the Government of Canada through the Department of Innovation, Science
and Economic Development and by the Province
of Ontario through the Ministry of Research, Innovation and Science. A.K. thanks Gerard Milburn and Sally Shrapnel for helpful
discussions. 

\bibliographystyle{unsrt}
\bibliography{library}

\begin{thebibliography}{10}

\bibitem{Kent2021b}
Adrian Kent.
\newblock Quantum state readout, collapses, probes, and signals.
\newblock {\em Physical Review D}, 103, 2021.

\bibitem{Page1981}
Don~N. Page and C.~D. Geilker.
\newblock Indirect evidence for quantum gravity.
\newblock {\em Physical Review Letters}, 47, 1981.

\bibitem{WilsonGerow2022}
Jordan Wilson-Gerow and Philip C.~E. Stamp.
\newblock Propagators in the correlated worldline theory of quantum gravity.
\newblock {\em Physical Review D}, 105, 2022.

\bibitem{Kuo1993}
Chung~I. Kuo and L.~H. Ford.
\newblock Semiclassical gravity theory and quantum fluctuations.
\newblock {\em Physical Review D}, 47, 1993.

\bibitem{Tilloy2016}
Antoine Tilloy and Lajos Diósi.
\newblock Sourcing semiclassical gravity from spontaneously localized quantum matter.
\newblock {\em Physical Review D}, 93, 2016.

\bibitem{tilloy_binding_2018}
Antoine Tilloy.
\newblock Binding {Quantum} {Matter} and {Space}-{Time}, {Without} {Romanticism}.
\newblock {\em Foundations of Physics}, 48(12):1753--1769, December 2018.

\bibitem{Gisin1990}
Nicolas Gisin.
\newblock Weinberg's non-linear quantum mechanics and supraluminal communications.
\newblock {\em Physics Letters A}, 143, 1990.

\bibitem{Kent2005}
Adrian Kent.
\newblock Nonlinearity without superluminality.
\newblock {\em Physical Review A - Atomic, Molecular, and Optical Physics}, 72, 2005.

\bibitem{Masanes2025}
Llu{\'{i}}s Masanes, Thomas~D. Galley, and Markus~P. M{\"{u}}ller.
\newblock Response to ``{T}he measurement postulates of quantum mechanics are not redundant".
\newblock {\em {Quantum}}, 9:1592, January 2025.

\bibitem{Kent2025}
Adrian Kent.
\newblock The measurement postulates of quantum mechanics are not redundant.
\newblock {\em {Quantum}}, 9:1749, May 2025.

\bibitem{Stacey2024}
Blake~C. Stacey.
\newblock Contradictions or curiosities? {O}n {K}ent's critique of the {Masanes--Galley--M\"uller} derivation of the quantum measurement postulates, 2024.
\newblock arXiv:2405.17733 [quant-ph].

\bibitem{Einstein1935}
A.~Einstein, B.~Podolsky, and N.~Rosen.
\newblock Can quantum-mechanical description of physical reality be considered complete?
\newblock {\em Phys. Rev.}, 47:777--780, May 1935.

\bibitem{wightman_theorie_1964}
A.~S. Wightman.
\newblock La théorie quantique locale et la théorie quantique des champs.
\newblock {\em Annales de l'institut Henri Poincaré. Section A, Physique Théorique}, 1(4):403--420, 1964.

\bibitem{Halvorson2006col}
Hans Halvorson and Michael Mueger.
\newblock Algebraic quantum field theory.
\newblock In J.~Butterfield and J.~Earman, editors, {\em Handbook of the philosophy of physics}. Kluwer Academic Publishers, 2006.

\bibitem{Bose2017}
Sougato Bose, Anupam Mazumdar, Gavin~W. Morley, Hendrik Ulbricht, Marko Toroš, Mauro Paternostro, Andrew~A. Geraci, Peter~F. Barker, M.~S. Kim, and Gerard Milburn.
\newblock Spin entanglement witness for quantum gravity.
\newblock {\em Physical Review Letters}, 119, 2017.

\bibitem{Marletto2017}
Chiara Marletto and Vlatko Vedral.
\newblock Gravitationally induced entanglement between two massive particles is sufficient evidence of quantum effects in gravity.
\newblock {\em Physical Review Letters}, 119, 2017.

\bibitem{Howl2021}
Richard Howl, Vlatko Vedral, Devang Naik, Marios Christodoulou, Carlo Rovelli, and Aditya Iyer.
\newblock {Non-Gaussianity as a Signature of a Quantum Theory of Gravity}.
\newblock {\em PRX Quantum}, 2, 2021.

\bibitem{Kent2021a}
Adrian Kent and Damián Pitalúa-García.
\newblock {Testing the nonclassicality of spacetime: What can we learn from Bell–Bose et al. -Marletto-Vedral experiments?}
\newblock {\em Physical Review D}, 104, 2021.

\bibitem{Kent2021c}
Adrian Kent.
\newblock Testing quantum gravity near measurement events.
\newblock {\em Physical Review D}, 103, 2021.

\bibitem{VonNeumann1955}
John von Neumann.
\newblock {\em Mathematical Foundations of Quantum Mechanics}.
\newblock Princeton University Press, 1955.

\bibitem{Peres1993}
Asher Peres.
\newblock {\em Quantum Theory: Concepts and Methods}.
\newblock Kluwer Academic, 1993.

\bibitem{Hanggi2013}
Esther Hänggi and Stephanie Wehner.
\newblock A violation of the uncertainty principle implies a violation of the second law of thermodynamics.
\newblock {\em Nature Communications}, 4, 2013.

\bibitem{Mielnik1979}
Bogdan Mielnik.
\newblock Mobility of nonlinear systems.
\newblock {\em Journal of Mathematical Physics}, 21, 1979.

\bibitem{Galley2023}
Thomas~D. Galley, Flaminia Giacomini, and John~H. Selby.
\newblock Any consistent coupling between classical gravity and quantum matter is fundamentally irreversible.
\newblock {\em {Quantum}}, 7:1142, October 2023.

\bibitem{Terno2024}
Daniel~R. Terno.
\newblock Classical-quantum hybrid models, 2024.
\newblock arXiv:2309.05014 [quant-ph].

\bibitem{Fedida2025}
Samuel Fedida and Adrian Kent.
\newblock Mixture equivalence principles and postquantum theories of gravity.
\newblock {\em Phys. Rev. D}, 111:126016, Jun 2025.

\bibitem{Susskind1993}
Leonard Susskind, Lárus Thorlacius, and John Uglum.
\newblock The stretched horizon and black hole complementarity.
\newblock {\em Physical Review D}, 48, 1993.

\bibitem{Krumm2017}
Marius Krumm, Howard Barnum, Jonathan Barrett, and Markus~P. Müller.
\newblock Thermodynamics and the structure of quantum theory.
\newblock {\em New Journal of Physics}, 19, 2017.

\bibitem{Layton2025}
Isaac Layton and Harry J.~D. Miller.
\newblock Restoring the second law to classical-quantum dynamics, 2025.

\bibitem{Jones1995}
Kingsley R.~W. Jones.
\newblock Newtonian quantum gravity.
\newblock {\em Australian Journal of Physics}, 48(6):1055, 1995.

\bibitem{Ruffini1969}
Remo Ruffini and Silvano Bonazzola.
\newblock Systems of self-gravitating particles in general relativity and the concept of an equation of state.
\newblock {\em Phys. Rev.}, 187:1767--1783, Nov 1969.

\bibitem{Barnum2007}
Howard Barnum, Jonathan Barrett, Matthew Leifer, and Alexander Wilce.
\newblock Generalized no-broadcasting theorem.
\newblock {\em Physical Review Letters}, 99, 2007.

\bibitem{Short2010}
Anthony~J. Short and Stephanie Wehner.
\newblock Entropy in general physical theories.
\newblock {\em New Journal of Physics}, 12, 2010.

\bibitem{Szilard1929}
Leo Szilard.
\newblock {über die Entropieverminderung in einem thermodynamischen System bei Eingriffen intelligenter Wesen}.
\newblock {\em Zeitschrift für Physik}, 53, 1929.

\bibitem{Deutsch1983}
David Deutsch.
\newblock Uncertainty in quantum measurements.
\newblock {\em Physical Review Letters}, 50, 1983.

\bibitem{Maassen1988}
Hans Maassen and J.~B.~M. Uffink.
\newblock Generalized entropic uncertainty relations.
\newblock {\em Phys. Rev. Lett.}, 60:1103--1106, Mar 1988.

\bibitem{Oppenheim2010}
Jonathan Oppenheim and Stephanie Wehner.
\newblock The uncertainty principle determines the nonlocality of quantum mechanics.
\newblock {\em Science}, 330, 2010.

\bibitem{Muller2012}
Markus~P. Müller, Jonathan Oppenheim, and Oscar~C.O. Dahlsten.
\newblock The black hole information problem beyond quantum theory.
\newblock {\em Journal of High Energy Physics}, 2012, 2012.

\end{thebibliography}

\end{document}